\documentclass{iopart} % omit 12pt to obtain normal journal page and type size

\usepackage[utf8]{inputenc}

\usepackage{iopams}
\usepackage{graphicx}
\graphicspath{{images/}}
\usepackage{xcolor}
\usepackage[colorlinks=true]{hyperref}
\usepackage{bbold}
\usepackage[numbers,sort&compress]{natbib}
\usepackage{placeins}

\newcommand{\au}{a.u.}
\newcommand{\bbone}{\ensuremath{\mathbb{1}}}
\newcommand{\bra}[1]{\langle{}#1|}
\newcommand{\braket}[2]{\langle{}#1|#2\rangle{}}
\newcommand{\ci}{\rmi} % the complex i
\newcommand{\diffd}{\rmd} % the differential d
\newcommand{\DIT}{DIT}
\newcommand{\eg}{{\it e.g.\ }}
\newcommand{\ie}{{\it i.e.\ }}
\newcommand{\ket}[1]{|#1\rangle{}}
\renewcommand{\Im}{\mathrm{Im}}
\newcommand{\matel}[3]{\langle{}#1|#2|#3\rangle{}}
\renewcommand{\Re}{\mathrm{Re}}
\newcommand{\SIT}{SIT}
\newcommand{\vecop}[1]{\boldsymbol{#1}}
\newcommand{\Zee}{\emph{Z\lowercase{ee}}}
\newcommand{\eZe}{\emph{\lowercase{e}Z\lowercase{e}}}
\begin{document}

\title[Extracting partial decay rates of helium from complex rotation]
{Extracting partial decay rates of helium from complex rotation: autoionizing
resonances of the one-dimensional configurations}

\author{Klaus~Zimmermann$^1$, Pierre~Lugan$^{1,2}$,
  Felix~J\"order$^1$, Nicolai~Heitz$^1$, Maximilian~Schmidt$^1$,
  Celsus~Bouri$^{1,3}$, Alberto~Rodriguez$^1$, and
  Andreas~Buchleitner$^1$}

\address{
$^1$ Physikalisches Institut, Albert-Ludwigs-Universit\"at Freiburg,
Hermann-Herder-Str.~3, D-79104 Freiburg, Germany\\
$^2$ Institute of Theoretical Physics, Ecole Polytechnique
F\'ed\'erale de Lausanne EPFL, CH-1015 Lausanne, Switzerland\\
$^3$ Centre of Atomic Molecular Physics and Quantum Optics (CEPAMOQ),
Faculty of Science, University of Douala, Douala, P.O. Box 8580,
Cameroon
}

\ead{klaus.zimmermann@physik.uni-freiburg.de}

\begin{abstract}
Partial autoionization rates of doubly excited one-dimensional helium
in the collinear \Zee{} and \eZe{} configuration are obtained by means
of the complex rotation method. The approach presented here relies on
a projection of \emph{back-rotated} resonance wave functions onto
singly ionized $\textrm{He}^{+}$ channel wave functions and the
computation of the corresponding particle fluxes. In spite of the
long-range nature of the Coulomb potential between the electrons and
the nucleus, an asymptotic region where the fluxes are stationary is
clearly observed. Low-lying doubly excited states are found to decay
predomintantly into the nearest single-ionization continuum. This
approach paves the way for a systematic analysis of the decay rates
observed in higher-dimensional models, and of the role of electronic
correlations and atomic structure in recent photoionization
experiments.
\end{abstract}

%\pacs{}
%\noindent{\it Keywords\/}:
%\maketitle

\section{Introduction\label{sec:introduction}}

With only three constituents the helium atom is one of the simplest prototypes
of a complex system in the realm of atomic physics, and it enjoys a long history
of theoretical and experimental studies~\cite{tanner_theory_2000}. The
complexity of helium is rooted in the long-range Coulomb interaction between the
electrons and the nucleus, which gives rise to a wealth of interesting phenomena
and challenging problems, related in particular to the mixed regular-chaotic
phase space at the classical level~\cite{richter_analysis_1990}, to high
spectral densities below the double-ionization
threshold~\cite{jiang_unreliability_2011}, to the autoionization of
doubly-excited states~\cite{fano_effects_1961, madden_new_1963,
pisharody_probing_2004, eiglsperger_spectral_2010}, or to the impact of
electronic correlations in photoionization
processes~\cite{walker_precision_1994, byun_scaling_2007,
becker_many-electron_2008, palacios_two-photon_2010,
jorder_interaction_2014}, to name a few essential aspects.

The main difficulty in the description of helium is the presence of the Coulomb
interaction between the electrons, whereby the three-body dynamics cannot simply
be modeled by two independent-electron problems~\cite{fano_correlations_1983}.
The comparable magnitude of the Coulomb interaction between all three
constituents also precludes simple perturbative
schemes~\cite{leopold_semiclassical_1980, feagin_molecular_1986,
briggs_aspects_1992}. The complexity of the problem had been recognized early on, in
the last century, when unsuccessful attempts were made to describe the spectrum
of helium on the basis of early semiclassical quantization techniques. The
electron-electron interaction lies at the heart of this failure, as it gives
rise to mixed regular-chaotic structures in the high-dimensional phase space
spanned by the motion of the electrons around the nucleus. Modern semiclassical
theory has provided a better understanding about the link between quantum
spectra and underlying classical dynamics \cite{leopold_semiclassical_1980,
gutzwiller_chaos_1990, ezra_semiclassical_1991, tanner_theory_2000,
choi_classical_2004}. In the case of helium, however, this link remains rather
formal, as only part of the helium spectrum is so far understood in
semiclassical terms. These limitations can be traced back to the fact that the
high-dimensional classical phase space of the helium atom remains itself in its
largest parts unexplored beyond the vicinity of one-dimensional configurations
and the triple-collision point~\cite{byun_scaling_2007,
  sano_periodic_2008}, but also more
fundamentally to the absence of an established semiclassical theory to describe
the decay rates of an open quantum system such as helium
\cite{keshavamurthy_dynamical_2011}.

At the level of the quantum spectrum of helium, the electron-electron
interaction has two nontrivial consequences. First, while the Rydberg series of
an independent-electron picture are scrambled, the excited states of helium are
still organized in series~\cite{burgers_highly_1995}. The understanding of the
symmetries underlying these spectral series and their classification according
to new quantum numbers constitute a remarkable achievement of the
group-theoretical~\cite{herrick_supermultiplet_1980, herrick_new_1983} and
adiabatic approaches~\cite{macek_properties_1968, fano_correlations_1983,
lin_classification_1984, lin_hyperspherical_1995, feagin_molecular_1986,
feagin_molecular-orbital_1988, rost_saddle_1991} developed by several authors
over the last three decades. Second, all doubly excited states of helium are in fact
resonance states, prone to various degrees of decay via autoionization
\cite{madden_new_1963, domke_high-resolution_1996,
eiglsperger_spectral_2010}. Qualitatively, the formation of these decaying
resonance states can be traced back to the Coulomb coupling of the doubly
excited bound states of the independent-electron picture to neighbouring
single-ionization continua. This picture, however, provides only little insight
into {\it how} doubly excited states of helium autoionize, as the
electron-electron coupling is nonperturbative. Unravelling the role of
electronic correlations in few-electron atoms is in fact of prime interest to
explain not only nonradiative decay via autoionization, but also photoionization
processes or the nature of chemical bonds, as shown by recent and intense
experimental and theoretical activity~(see
Refs.~\cite{fittinghoff_observation_1992, walker_precision_1994,
palacios_two-photon_2010, ott_quantum_2012} and references therein).

In the framework of the molecular adiabatic
approach~\cite{feagin_molecular_1986}, the decay of doubly excited two-electron
states can be understood as resulting from nonadiabatic transitions between
adiabatic potential curves~\cite{tanner_theory_2000,
feagin_molecular-orbital_1988, rost_propensity_1990, rost_resonance_1997}. To
analyze the mechanism of such couplings, a set of propensity rules was
established~\cite{rost_propensity_1990, rost_nodal_1991, rost_saddle_1991}.
These propensity rules are approximate selection rules that account for the
strong dominance of certain decay channels, based on the symmetries of the
states described by the molecular adiabatic classification. While these
rules were found to describe well the decay of resonances in
low-lying series, their predictive power is reduced for higher excitations, as
resonance series start to overlap and approximate quantum numbers are
progressively lost for principal quantum numbers $N\geq
6$~\cite{burgers_highly_1995, rost_resonance_1997}. For such regimes where
several decay channels may play comparable roles, resorting to tools allowing
the quantitative analysis of {\it partial decay rates} of resonance states (\ie
branching ratios between decay channels) from first principles appears
necessary. This need arises not only for autoionization processes, but also for
the analysis of electron-impact ionization and photoionization.

Various numerical {\it ab initio} approaches have been developed and intensively
used over the past thirty years to study the consequences of electron-electron
correlations on spectra and ionization processes in helium (for a review see
Refs.~\cite{lambropoulos_two-electron_1998, tanner_theory_2000} and references
therein). Among those, the method of complex rotation~\cite{aguilar_class_1971,
balslev_spectral_1971, simon_quadratic_1972, simon_resonances_1973, simon_resonances_1978, reed_methods_1978, reinhardt_complex_1982,
junker_recent_1982, ho_method_1983, lindroth_calculation_1994,
moiseyev_quantum_1998, buchleitner_microwave_1995,
buchleitner_wavefunctions_1994} has established itself as a powerful
technique allowing
the computation of resonance positions and widths as well as ionization cross
sections with unprecedented accuracy, including high-lying resonance
series~\cite{wintgen_double_1993, burgers_highly_1995,
gremaud_photo-ionization_1997, madronero_decay_2005, eiglsperger_highly_2009,
eiglsperger_spectral_2010}. However, in contrast to time-dependent
calculations~\cite{scrinzi_ionization_1997, foumouo_theory_2006,
nikolopoulos_time-dependent_2007, hochstuhl_two-photon_2011} and other
time-independent numerical approaches relying \eg on $R$-matrix
techniques~\cite{burke_R-matrix_1975, hayes_resonances_1988,
malegat_double_1999, selles_ab_2002}, the Feshbach projection
formalism~\cite{sanchez_extensive_1993}, or close coupling
methods~\cite{tang_general_1992, tang_evidence_1992, tang_mechanism_1994,
lin_hyperspherical_1995, bray_electrons_2002}, which have been used successfully
to compute partial ionization rates in regimes of moderate
excitation~\cite{menzel_decay_1995, menzel_partial_1996} and close to the
three-body breakup threshold~\cite{bouri_one-photon_2007}, the application of
the complex-rotation technique to helium has hitherto mostly been restricted to
the calculation of {\it total} decay rates and cross
sections~\cite{tanner_theory_2000}. A notable exception to this state of affairs
for the three-body Coulomb problem is provided by a calculation of
angle-resolved differential cross sections for the electron-impact ionization of
hydrogen~\cite{rescigno_collisional_1999}.

Early schemes to calculate partial decay rates within the framework of complex
rotation have been devised in Refs.~\cite{noro_resonance_1980,
bacic_complex_1982, peskin_partial_1990, moiseyev_partial_1990}. The key element
in these proposals is that, while the complex eigenenergies obtained from
complex rotation yields only the resonance positions and widths, the associated
resonance wave functions (eigenvectors) may be used to compute transitions
matrix elements~\cite{noro_resonance_1980}, or projected onto channel wave
functions~\cite{bacic_complex_1982, peskin_partial_1990, moiseyev_partial_1990}
to deduce the associated partial decay rates. The above schemes rely on the use
of the (square-integrable) complex-rotated resonance wave function, and require
either an integration over the radial degree of freedom along which decay
occurs~\cite{noro_resonance_1980, bacic_complex_1982} or plane-wave asymptotics
of the resonance wave function~\cite{peskin_partial_1990,
moiseyev_partial_1990}, which precludes a straightforward application to
long-ranged Coulomb potentials.

We employ here an alternative approach to calculate partial decay rates, which
was initiated in Ref.~\cite{schmidt_partial_2009} (a related approach
was examined in Ref.~\cite{goldzak_evaluation_2010}) and
which relies on an inspection of the particles fluxes associated with the {\it
back-rotated} resonance wave functions. With this method, we analyze the partial
decay rates of atomic resonance states in two one-dimensional models
of helium, namely the so-called \Zee{} and \eZe{} configurations, where the two
electrons lie on the same or on opposite sides of the nucleus,
respectively. Such collinear models of helium lack the angular degrees of
freedom, but contain essential ingredients of the full-dimensional helium atom,
such as autoionizing resonance states, overlapping series in the highly doubly
excited regime, high spectral densities and an accumulation of series close to
the double-ionization threshold. Moreover, these configurations correspond
to invariant subspaces of classical motion in 3D
helium~\cite{tanner_theory_2000}. The \Zee{} subspace and its vicinity
have been
analyzed thoroughly~(see \eg Refs.~\cite{ostrovsky_planetary_1995,
schlagheck_classical_1998}) as it has been recognized, in particular, that they
host pronounced resonances associated with stable islands in classical phase
space, known as {\it frozen-planet configurations}~\cite{richter_stable_1990}.
Our approach for the calculation of partial decay rates may thus be useful not
only for the quantitative analysis of fragmentation processes in strongly
correlated few-electron systems, but also from the point of view of the
semiclassical physics of mixed phase spaces~\cite{keshavamurthy_dynamical_2011}.

This article is organized as follows. In section~\ref{sec:Zee_wavefunctions} we
briefly review the spectral structure of collinear helium, and we present
the analytical and numerical tools used for the description of resonance states
within the framework of complex rotation. In section~\ref{sec:partial_rates} we
present our general scheme for the calculation of partial decay rates. In
section~\ref{sec:partial_rates_Zee} this procedure is used to analyze the
decay of autoionizing doubly excited states of \Zee{} and \eZe{} helium. In
section~\ref{sec:conclusion}, ultimately, we summarize our results.

\section{Wave functions of atomic resonance states in one-dimensional
  helium\label{sec:Zee_wavefunctions}}

In the one-dimensional (1D) model of helium, the nucleus and the two
electrons are constrained to move on one fixed line, with the
electrons either both residing on one side of the nucleus (\Zee{}), or
on opposite sides (\eZe{}). In the limit of an infinitely massive
nucleus, and neglecting relativistic effects, the dynamics of the
electrons with respect to the nucleus is governed by the Hamiltonian
%%%%
\begin{equation} \label{eq:Hzz}
H=-\frac{1}{2}\frac{\partial^2}{\partial
z_1^2}-\frac{1}{2}\frac{\partial^2}{\partial
z_2^2}-\frac{Z}{z_1}-\frac{Z}{z_2}+\frac{\gamma}{|z_2 \mp z_1|},
\end{equation}
%%%%
where $z_i>0$ is the distance of electron $i$ from the nucleus, $Z=2$ is the
Coulomb charge of the nucleus, $\gamma$ is a numerical parameter accounting for
the strength of Coulomb interaction between the electrons ($\gamma=1$ amounts to
the natural interaction), and Hartree atomic units ($\textrm{\au}$) have been
used: $e \equiv m_e \equiv \hbar \equiv 1/(4\pi \epsilon_0) \equiv 1$. With the
latter, the length unit is the Bohr radius $a_0$ and the energy unit is the
Hartree ($\sim 27.2~\textrm{eV}$)~\cite{baylis_units_2006}.
The Hamiltonian $H$ acts on two-body wave functions $\psi(z_1, z_2)$
that are either symmetric (even) or antisymmetric (odd) under the
exchange of coordinates. In the \Zee{} case this symmetry originates from
particle exchange, and corresponds to singlet and triplet spin states,
respectively. In the \eZe{} case, the symmetry arises from the invariance
of $H$ under the combined action of particle exchange and central
symmetry, i.e. reflection of all coordinates about the
origin. Furthermore, irrespective of symmetry, the Coulomb potential
between electrons demands a nodal line on the $z_1=z_2$ line of \Zee{}
configurations, effectively rendering the even and odd wave functions
degenerate and identical, save for a sign in half of configuration
space. As a consequence the configuration space may be reduced to the
sector $z_1>z_2$, and we will hereafter simply talk about \Zee{} helium
without distinction of symmetry. In contrast, symmetric \eZe{} states
show an antinode on the $z_1=z_2$ line, so that the \eZe{} states need to
be distinguished according to symmetry. This symmetry has important
repercussions on the physical properties of eigenstates as, in
particular, the nodal $z_1=z_2$ line of odd \eZe{} states leads to a
vanishing density close to the triple collision point and consequently
to an increased stability compared to the even states
\cite{richter_classical_1993}.

\subsection{Autoionization resonance states in
helium\label{subsec:resonance_states}}

\begin{figure}
    \includegraphics{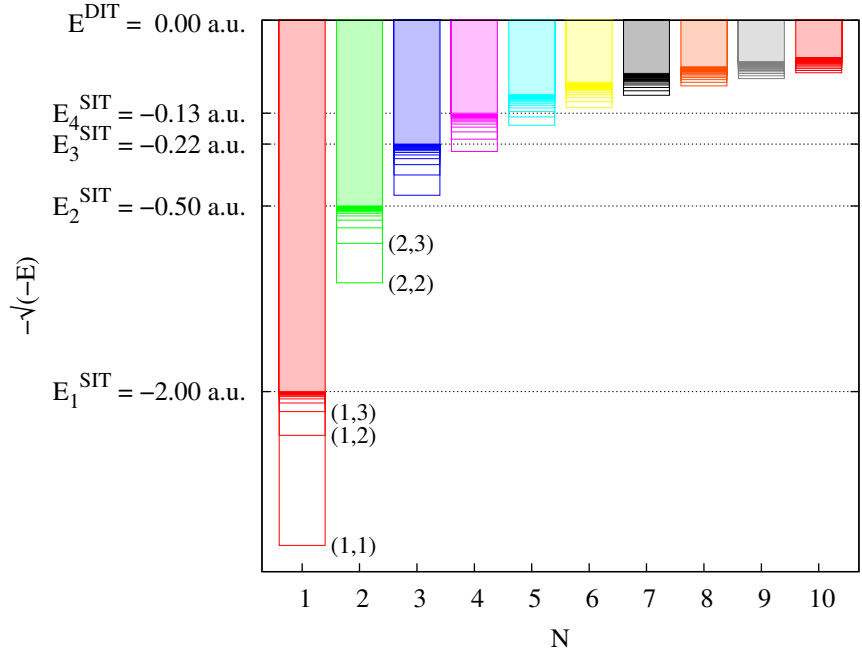}
    \caption{\label{fig:he-spectrum}
      (Colour online) Schematic spectrum of 1D helium without
      electron-electron interaction. The quantum numbers of the two
      electrons are $N$ and $n$.
      For every fixed $N$ there is a Rydberg series with $n>N$ ($n\geq
      N$ for even states) that converges to the respective
      single-ionisation threshold. Note that in the case of even
      (singlet) Zee configurations, the $n=N$ states disappear as soon
      as even a weak Coulomb potential is turned on between the
      electrons.
      To exemplify the notation, we explicitly label five states by
      $(N,n)$.
      The single-ionization thresholds of the first four
      series are marked. Note that the ground state of the second
      series is energetically already embedded in the continuum of the
      first. Note further, that in this uncoupled picture the
      series start to overlap already from $N=5$ onwards, \ie the ground
      state of the fifth series is lower in energy, than the
      single-ionization threshold of the fourth. The single-ionization
      thresholds $E_N^\mathrm{\SIT}=-2/N^2~\textrm{\au}$ and the
      double-ionization threshold $E^\mathrm{\DIT}=0~\textrm{\au}$ are left
      unaffected by a non-vanishing electron-electron coupling.
    }
\end{figure}

In the absence of interaction between the electrons ($\gamma=0$), the
electronic dynamics is separable, and $H$ reduces to the sum of two
hydrogen-like Hamiltonians, each of which gives rise to a Rydberg
series of bound states and a continuum of Coulomb scattering
states. The resulting two-electron spectrum is depicted in
Fig.~\ref{fig:he-spectrum}. Each bound state of the lowest-lying
electron, labelled by the quantum number $N$, comes along with a
Rydberg series $E_{N n}= E_{N}^{\mathrm{\SIT}}
-Z^2/(2n^2)~\textrm{\au}$, where $n\geq N$ (even, singlet) or $n>N$
(odd, triplet), and a continuum of singly-ionized helium states above
the $N$-th single-ionization threshold $E_{N}^{\mathrm{\SIT}}=
-Z^2/(2N^2)~\textrm{\au}$ For increasing $N$, these series converge to
the double-ionization threshold $E^{\mathrm{\DIT}}=0~\textrm{\au}$

The inclusion of the Coulomb interaction between the electrons does
not change the positions of the ionization thresholds, but strongly
affects the discrete part of the spectrum. The discrete levels of the
Rydberg series couple among themselves and to the ionization
continua. As a result, all levels are shifted and the continuum
coupling turns doubly excited bound states with $n\geq N > 1$ into
resonance states with a finite lifetime. Doubly excited states of 1D
helium thus decay naturally via autoionization into singly ionized
$\textrm{He}^{+}$ ions, without any perturbation by an external
field. This phenomenology also holds in higher
dimensions~\cite{tanner_theory_2000, fano_effects_1961}. In
particular, the \Zee{} autoionization resonances discussed here arise
as the 1D counterparts of the frozen-planet states known from
higher-dimensional models of helium~\cite{richter_stable_1990,
richter_calculations_1991}, although the lifetimes of the latter are
considerably shorter, due to the allowed excursions from
collinearity~\cite{schlagheck_drei-korper-coulombproblem_1999,
madronero_decay_2005}. The \eZe{} resonances, in contrast, exhibit
much larger decay rates, which is not surprising in the light of the fully
chaotic classical phase space of this configuration
\cite{ezra_semiclassical_1991}. Yet, while doubly excited states
decay, they may be long-lived enough and their density may be such
that they play an important role in the internal structure of the
atom, leaving strong signatures in scattering and photoionization
signals~\cite{madden_new_1963, brotton_electron-impact_1997,
  jiang_unreliability_2011, ott_quantum_2012}.

%%%%
%%%%

While the nonperturbative Coulomb coupling invalidates a
straightforward classification of the resonance states of helium in
terms of the independent-electron picture, these resonances remain
organized in well-identifiable series that converge to
the different single-ionization thresholds, for low excitation (see
Fig.~\ref{fig:series_4_spectrum}). In 1D \Zee{} helium, the bound
states below the first single-ionization threshold and the low-lying
resonances may be labelled by approximate quantum numbers $(N,n)$,
where $N$ denotes the series converging to the $N$-th
single-ionization threshold, and $n$ corresponds to the degree of
excitation within that series.  While the electrons are
indistinguishable, $N$ and $n$ may be seen as describing the state of
the {\it inner} and the {\it outer} electron, respectively, in analogy
with the uncoupled case ($\gamma=0$). This description is expected to
be valid at least up to $N=20$, where series start to
overlap~\cite{schlagheck_nondispersive_2003}. For \eZe{} also the
symmetry, even or odd, must be stated in addition
to the approximate quantum number $(N,n)$.

\subsection{Complex rotation\label{subsec:complex_rotation}}

Analyzing the formation of autoionizing resonance states in helium beyond the
qualitative picture given above is challenging for two reasons. First, the
strong Coulomb coupling of the electrons excludes perturbative approaches which
would be based on the independent-electron picture. For instance, while standard
Feynman-Dyson perturbation theory formally allows an exact description of the
continuum coupling, the complex internal structure of the atom depicted above
complicates the identification of a relevant subset of diagrams in most
situations (see \eg Ref.~\cite{hochstuhl_nonequilibrium_2010} for a discussion
in the framework of nonequilibrium Green functions). Second, resonance states
are not stationary, and their representation on a basis of stationary states of
either the uncoupled ($\gamma=0$) or the coupled Hamiltonian ($\gamma>0$)
necessarily involves a continuum of states~\cite{fano_effects_1961}. The
technique of complex rotation~\cite{moiseyev_quantum_1998} circumvents both
of these difficulties, as it allows the description of resonance states with
bound-state techniques and numerical calculations with sets of integrable
functions~\cite{doolen_electron-hydrogen_1974}, as outlined below.

The foundations and the applications of the complex-rotation method have been
reviewed \eg in Refs.~\cite{simon_resonances_1978, junker_recent_1982,
reinhardt_complex_1982, ho_method_1983, moiseyev_quantum_1998,
buchleitner_wavefunctions_1994}. In brief, Hamiltonian $H$ is replaced
by
%%%%
\begin{equation}\label{eq:Htheta}
H_\theta=R(\theta) H R(-\theta),
\end{equation}
%%%%
where $\theta$ is a real parameter, and $R(\theta)$ is the
complex-rotation operator defined by
%%%%
\begin{equation}\label{eq:Rtheta}
R(\theta)=\exp\left(
-\theta\frac{\vecop{r}.\vecop{p}+\vecop{p}.\vecop{r}}{2}
\right),
\end{equation}
%%%%
where $\vecop{r}$ and $\vecop{p}$ are position and momentum operators for a
$d$-dimensional configuration space. While the parameter of the complex-rotation
operator $R$ itself may be positive or negative, the $\theta$ parameter entering
equation~(\ref{eq:Htheta}) is typically taken as a small positive angle for
resonance poles belonging to the \textit{lower} half complex plane to appear in
the resolvent of the operator $H_\theta$ (see below). Practically, the rotated
Hamiltonian $H(\theta)$ is obtained from $H$ through the substitution of
$\vecop{r}$ and $\vecop{p}$ by $\vecop{r}_\theta= R(\theta) \vecop{r}
R(-\theta)= \vecop{r} \, e^{\ci\theta}$ and $\vecop{p}_\theta= R(\theta)
\vecop{p} R(-\theta)= \vecop{p} \, e^{-\ci\theta}$. In particular, this complex
scaling of coordinates turns Hamiltonian~(\ref{eq:Hzz}) into
%%%%
\begin{equation} \label{eq:Hzztheta}
H_\theta=
-\frac{1}{2}e^{-2\ci\theta}\left(
\frac{\partial^2}{\partial z_1^2}
+\frac{\partial^2}{\partial z_2^2}
\right)
+e^{-\ci\theta}\left(
-\frac{Z}{z_1}-\frac{Z}{z_2}+\frac{\gamma}{|z_2 \mp z_1|}
\right).
\end{equation}
%%%%
For real, nonzero $\theta$, the operator $H_\theta$ is no longer Hermitian on
the $L^2$ Hermitian domain of $H$, but becomes complex symmetric:
$H_\theta^\dagger=\overline{H_\theta}=H_{-\theta}$, where the bar denotes
complex conjugation. As $H_\theta$ is not Hermitian, its eigenvalues are not
restricted to the real axis, and the eigenvectors will in general not be their
own biorthogonal adjoints~\cite{morse_methods_1953}. Because of the complex
symmetry, however, the eigenvectors $\ket{\psi_i^\theta}$ of $H_\theta$ satisfy
the following biorthogonality relation:
%%%%
\begin{equation} \label{eq:biorthogonality}
\braket{\overline{\psi_i^\theta}}{\psi_j^\theta}
= \int \diffd\boldsymbol{r}\,
\psi_i^\theta (\boldsymbol{r}) \psi_j^\theta (\boldsymbol{r})
= \delta_{ij},
\end{equation}
%%%%
where $\braket{.}{.}$ is the usual Hermitian product, the bar again denotes
complex conjugation, and the Kronecker delta may be replaced by a delta
distribution for continuum states.

The spectrum of $H$ is defined as the set of complex numbers $z$ for which $H$
admits no bounded resolvent operator (Green's function)
$G(z)=(z-H)^{-1}$~\cite{reed_methods_1972}. As outlined in
section~\ref{subsec:resonance_states}, this spectrum contains distinct proper
eigenvalues, which show up as isolated poles in the resolvent $G(z)$ and
correspond to bound states of the atom, as well as continua of improper
eigenvalues, which lead to branch cuts in $G(z)$ and correspond to scattering
states. Under complex rotation, this spectrum transforms as follows:
%%%%
\begin{enumerate}
\item The proper eigenvalues of $H_\theta$ coincide with those of $H$.
\item The continua of $H_\theta$ correspond to the continua of $H$, but are
rotated about the ionization thresholds downwards ($\theta>0$) into the complex
plane by an angle of $2\theta$.
\item For sufficiently large $\theta$, isolated complex eigenvalues
$E_i^{\theta} = E_i - \ci\Gamma_i/2$ appear in the lower complex half plane
($\Gamma_i>0$). In the limit of narrow and isolated
resonances~\cite{moiseyev_non-hermitian_2011}, these eigenvalues correspond to
individual resonance states of $H$ with position $E_i$ and decay rate
$\Gamma_i$, as discussed below.
\end{enumerate}
%%%%
The eigenstates $\ket{\psi_i^\theta}$ pertaining to the isolated complex
eigenvalues of $H_\theta$ are square-integrable. The states $\ket{\psi_i} =
R(-\theta)\ket{\psi_i^\theta}$ obtained upon \textit{back-rotation} are
eigenstates of $H$, with the same complex eigenvalue $E_i^\theta$, and they
decay under the action of $U(t)=\exp(-\ci H t)$ at rate $\Gamma_i$. The
\textit{resonance wave functions} associated with these back-rotated states
typically diverge exponentially as a function of distance, and thus lie outside
the Hermitian domain of $H$. As a matter of fact, they may be found by imposing
purely outgoing boundary conditions on the differential operator
$H$~\cite{moiseyev_non-hermitian_2011}, but such an approach may prove difficult
in practice~\cite{ho_method_1983}. In the setting of complex rotation, on the
other hand, the square-integrable rotated resonance states $\ket{\psi_i^\theta}$
can be accurately calculated on a large but finite $L^2$ basis set. Let us now
examine how these states may be used to describe the correlated electron
dynamics.

\subsection{Complex-rotated states and real atomic dynamics}

While the eigenstates of $H_\theta$ themselves are unphysical, they provide a
useful representation to describe the evolution of physical states under the
action of the Hamiltonian $H$. Under the assumption that their set is complete, the
eigenstates $\ket{\psi_i^\theta}$ of $H_\theta$ yield the resolution of identity
%%%%
\begin{equation} \label{eq:resolution_identity}
\sum_i \ket{\psi_i^\theta}\bra{\overline{\psi_i^\theta}}=\bbone,
\end{equation}
%%%%
where the sum runs over the bound states, the $L^2$ resonance states and the
continuum states of $H_\theta$. The time evolution of a square integrable state
$\ket{\phi(t)} = U(t)\ket{\phi(0)}$ and its overlap with a localized
time-independent reference state $\ket{\phi_0}$ are then given
by~\cite{maquet_stark_1983, buchleitner_microwave_1995}
%%%%
\begin{eqnarray} \label{eq:time_evo}
\braket{\phi_0}{\phi(t)}
&=\matel{\phi_0}{R(-\theta)e^{-\ci H_\theta t}R(\theta)}{\phi(0)}\nonumber\\
&=\sum_i e^{-\ci E_i^{\theta} t}
\matel{\phi_0}{R(-\theta)}{\psi_i^\theta}
\matel{\overline{\psi_i^\theta}}{R(\theta)}{\phi(0)}.
\end{eqnarray}
%%%%
The initial state $\ket{\phi(0)}$ may be a nonstationary state of $H$ obtained
\eg by applying an external field to the atom in its ground state. When
$\ket{\phi(0)}$ overlaps mostly with a single (back-rotated) resonance wave
function $\ket{\psi_{i_0}} = R(-\theta)\ket{\psi_{i_0}^\theta}$, the dominant
feature in the time evolution of $|\braket{\phi_0}{\phi(t)}|^2$ is an
exponential decay at rate $\Gamma_{i_0} = -2\,\Im(E_i^\theta)$ on intermediate time scales~\cite{simon_resonances_1973}.

The above considerations can be developed in a more general setting and couched
in the resolvent formalism. Indeed, identity~(\ref{eq:resolution_identity})
allows for a spectral representation of the Green's function of $H$ in terms of
the eigenstates of $H_\theta$~\cite{johnson_l^2_1983}. More precisely, complex
rotation directly provides the analytical continuation $G^{II}(z)$ of
$G(z)=(z-H)^{-1}$ on the second Riemann
sheet~\cite{buchleitner_wavefunctions_1994}\footnote{See also Ref.~\cite{cohen-tannoudji_atom_2008} for a general discussion of resonance poles and analytic continuation without the framework complex rotation.}:
%%%%
\begin{equation} \label{eq:GIIz}
G^{II}(z)
= R(-\theta)\frac{1}{z-H_\theta}R(\theta)
=\sum_i\frac{
R(-\theta)\ket{\psi_i^\theta}\bra{\overline{\psi_i^\theta}}R(\theta)
}{z-E_i^\theta}.
\end{equation}
%%%%
This continuation is sound when expression~(\ref{eq:GIIz}) is applied on both
sides on square integrable states, and $R(-\theta)$ and $R(\theta)$ are
interpreted as acting to the left and to the right,
respectively~\cite{buchleitner_wavefunctions_1994}. Expression~(\ref{eq:GIIz})
is free of branch cuts on the real axis. For real~$E$ and $\theta>0$ (resp.
$\theta<0$), $G^{II}(E)$ thus coincides with the retarded (resp. advanced)
Green's function $G^{\pm}(E) = \lim_{\eta\to 0^{+}} G(E\pm\ci \eta)$. Therefore,
the difference between the retarded and advanced Green's functions may be cast
into the form~\cite{buchleitner_wavefunctions_1994}
%%%%
\begin{equation} \label{eq:GpmGm}
\hspace{-1cm}
G^+(E)-G^-(E)=\sum_i \left[
  \frac{
R(-\theta)\ket{\psi_i^\theta}\bra{\overline{\psi_i^\theta}}R(\theta)
}{E-E_i^\theta}
-\frac{
R(\theta)\ket{\overline{\psi_i^\theta}}\bra{\psi_i^\theta}R(-\theta)
}{E-\overline{E_i^\theta}}
\right],
\end{equation}
%%%%
with $\theta>0$. Note that this expression also
writes~\cite{cohen-tannoudji_atom_2008}
%%%%
\begin{equation} \label{eq:projE}
G^+(E)-G^-(E)=-2\pi\ci \sum_j \rho_j(E) \ket{\psi_{E,j}}\bra{\psi_{E,j}},
\end{equation}
%%%%
where the $\ket{\psi_{E,j}}$ denote the eigenstates of $H$ with energy $E$, the
index $j$ labels other quantum numbers, and $\rho_j(E)$ is the associated
density of states. As a consequence of expression~(\ref{eq:GpmGm}), in
particular, the local density of states or \emph{electronic density at
energy}~$E$ is given by~\cite{buchleitner_wavefunctions_1994}
%%%%
\begin{eqnarray} \label{eq:densE}
\rho(E,\boldsymbol{r})
&=-\frac{1}{2\pi\ci}\left[
\matel{\boldsymbol{r}}{G^{+}(E)}{\boldsymbol{r}}
-\matel{\boldsymbol{r}}{G^{-}(E)}{\boldsymbol{r}}
\right]\nonumber\\
&=-\frac{1}{\pi}\Im
\matel{\boldsymbol{r}}{G^{+}(E)}{\boldsymbol{r}}
=\frac{1}{\pi}\Im
\sum_i\frac{\matel{\boldsymbol{r}}{R(-\theta)}{\psi_i^\theta}^2}{E_i^\theta-E}.
\end{eqnarray}
%%%%
Unlike expression~(\ref{eq:projE}), formulas (\ref{eq:GpmGm}) and
(\ref{eq:densE}) manifestly reflect the contributions of resonance states
(bound states diluted in a continuum) to the density of states. In the framework
of complex rotation, these contributions may be calculated accurately with
bound-state techniques. In the case where the sum~(\ref{eq:densE}) is restricted
to the contribution of one \textit{isolated} resonance pole $E_{i_0}^\theta$
($\Im\, E_{i_0}^\theta<0$), the associated electronic density reads
%%%%
\begin{equation} \label{eq:densEonepole}
\rho_{i_0}(\boldsymbol{r})
=\int \frac{\diffd E}{\pi} \Im\frac{\matel{\boldsymbol{r}}{R(-\theta)}{\psi_{i_0}^\theta}^2}{E_{i_0}^\theta-E}=\Re\left[\matel{\boldsymbol{r}}{R(-\theta)}{\psi_{i_0}^\theta}^2\right],
\end{equation}
%%%%
which typically diverges exponentially in space, as remarked above. The fact
that $\rho_{i_0}(\boldsymbol{r})$ may assume negative values locally can be
traced back to the single-pole approximation, as the sum over all states of
equation~(\ref{eq:densE}) would restore the positivity of the
density~\cite{buchleitner_wavefunctions_1994}. Note also that the difference
between
$\rho_{i_0}(\boldsymbol{r})=\Re[\psi_{i_0}(\boldsymbol{r})^2]=\Re[\braket{\boldsymbol{r}}{\psi_{i_0}}\braket{\overline{\psi_{i_0}}}{\boldsymbol{r}}]$
and
$|\matel{\boldsymbol{r}}{R(-\theta)}{\psi_{i_0}^\theta}|^2=|\psi_{i_0}(\boldsymbol{r})|^2=\braket{\boldsymbol{r}}{\psi_{i_0}}\braket{\psi_{i_0}}{\boldsymbol{r}}$
arises
from two ways of envisioning the single-pole approximation. The first one
involves a coupling from \eg a stationary state to the continuum at (real)
energy $E$ close to $E_{i_0}$, with a density of states given by
equation~(\ref{eq:GpmGm}), while the second one assumes that the physical state
is approximated by the back-rotated resonance wavefunction $\psi_{i_0}$ from
the outset.

Let us now revisit the situation described below equation~(\ref{eq:time_evo})
and consider the perturbation of Hamiltonian $H$ by an additional term $V'$, so
that the total Hamiltonian reads $H'=H+V'$. Then, the transition amplitudes
between stationary states $\ket{\phi_{in}}$ and $\ket{\phi_f}$ of $H$ in the
interaction picture are entirely determined by the elements of the
$T$-matrix~\cite{cohen-tannoudji_atom_2008},
%%%%
\begin{equation}
\matel{\phi_f}{T(E)}{\phi_{in}}=\matel{\phi_f}{V'+V' {G'}^{+}(E) V'}{\phi_{in}},
\end{equation}
%%%%
where ${G'}^{+}$ is the retarded Green's function of the total Hamiltonian
$H'$.In the limit of infinite perturbation times $\ket{\phi_{in}}$ couples only
to states $\ket{\phi_f}$ at energy $E_f=E_{in}$, and the transition rate between
$\ket{\phi_{in}}$ and a continuum of states $\ket{\psi_{E_{in},j}}$ at this
energy is
%%%%
\begin{eqnarray}
\gamma
&=2\pi \rho(E_{in})\sum_j |\matel{\psi_{E_{in},j}}{T(E_{in})}{\phi_{in}}|^2
\nonumber\\
&=\ci \bra{\phi_{in}} T^\dagger(E_{in})
\left[G^+(E_{in})-G^-(E_{in}) \right]
T(E_{in}) \ket{\phi_{in}}.
\end{eqnarray}
%%%%
In the Born approximation, $T(E)$ reduces to $V'$. Then, with the help of
equation~(\ref{eq:GpmGm}), the transition rate writes
%%%%
\begin{equation} \label{eq:gammaBorn}
\gamma
=2\, \Im \sum_i \frac{
\matel{\overline{\psi_i^\theta}}{R(\theta)V'}{\phi_{in}}^2
}{E_i^\theta-E_{in}}
=2\, \Im \sum_i \frac{
\matel{\phi_{in}}{V'R(-\theta)}{\psi_i^\theta}^2
}{E_i^\theta-E_{in}}.
\end{equation}
%%%%
As above, the initial state $\ket{\phi_{in}}$ may be the ground state $\ket{g}$
of the atom in the presence of a laser field with frequency $\omega$. In that case $H'$ involves the \textit{time-dependent} perturbation $V'(t)=-(\boldsymbol{D}^{+}+\boldsymbol{D}^{-}).(\boldsymbol{\mathcal{E}}^+ e^{-\ci \omega t}+\boldsymbol{\mathcal{E}}^- e^{\ci \omega t})$ in the dipole approximation, with $\boldsymbol{D}^{\pm}$ the
dipole transition operators and $\boldsymbol{\mathcal{E}}^\pm$ the classical field amplitudes~\cite{dalibard_atomic_1985}.
Then, upon substitution of $V'$ by $-\boldsymbol{D}^+.\boldsymbol{\mathcal{E}}^+$ and of $E_{in}$ by $E_g+\omega$, expression~(\ref{eq:gammaBorn}) directly gives the photoionization cross-section
$\sigma(\omega)$, at leading order in the field intensity~\cite{rescigno_rigorous_1975, buchleitner_wavefunctions_1994}. If
$V'\ket{\phi_{in}}$ overlaps mostly with the back-rotated vector $R(-\theta)
\ket{\psi_{i_0}^\theta}$ pertaining to an autoionizing resonance and $E_{in}$ is
chosen such that $|E_{in}-\Re(E_{i_0}^\theta)| \ll |\Im(E_{i_0}^\theta)|$,
expression~(\ref{eq:gammaBorn}) may be approximated by
%%%%
\begin{equation}
\gamma\simeq
\frac{4}{\Gamma_{i_0}} \,\Re\left(
\matel{\overline{\psi_{i_0}^\theta}}{R(\theta)V'}{\phi_{in}}^2
\right),
\end{equation}
%%%%
where $\Gamma_{i_0} = -2\,\Im(E_{i_0}^\theta)$ is the decay rate of the
resonance state. The fact that $\gamma$ is \textit{inversely} proportional to
$\Gamma_{i_0}$ is well known from resonance-mediated
decay~\cite{cohen-tannoudji_atom_2008}, and reflects that in the above example the
photoionization of $\ket{\phi_{in}}$ is mediated by the autoionizing resonance.
Most importantly, the back-rotated resonance wave function
$\psi_{i_0}(\boldsymbol{r}) =
\matel{\boldsymbol{r}}{R(-\theta)}{\psi_{i_0}^\theta}$ determines not only the
strength of the coupling between the initial state and the resonance state, but
also how the initial state decays into continuum states. Indeed, as emphasized
in the following sections, the resonance wave function carries in itself the
final (continuum) states of the decay process.

The above discussion illustrates that while the exact ionization dynamics of an
atomic state generally involves a continuum of eigenstates of $H_\theta$, the
most significant part of it may often be captured in a single resonance state
$\ket{\psi_{i_0}^\theta}$ under suitable conditions. The decay processes of such
individual resonance states is the subject of the remainder of this paper. We
emphasize that, rather than describing the transition from a stationary state to
the continuum in the vicinity of a resonance, as illustrated above, we shall
concentrate for simplicity on the assumption that the physical state under
scrutiny is described from the outset by the back-rotated wavefunction of that
resonance.

\subsection{Energies and wave functions of atomic resonances in 1D helium}
\label{subsec:numerics}

We use a {\em Sturmian basis} to compute the eigenvalues and
eigenvectors of the rotated Hamiltonian (\ref{eq:Hzztheta})
efficiently with standard techniques for sparse matrix eigenvalue
problems~\cite{coolidge_convergence_1937, pekeris_ground_1958,
  schlagheck_nondispersive_2003,
  schlagheck_drei-korper-coulombproblem_1999, wintgen_double_1993}.

Here the Sturmian functions are defined
by~\cite{schlagheck_nondispersive_2003}
%%%%
\begin{equation} \label{eq:1Dsturmians}
S_n^{(\alpha)}(r)=
\frac{1}{\sqrt{n}}(-1)^n\frac{2r}{\alpha}e^{-r/\alpha}
L_{n-1}^{(1)}\left(\frac{2r}{\alpha}\right),
\end{equation}
%%%%
where $L_{n}^{(1)}$ denotes an associated Laguerre polynomial of order
$n$~\cite{abramowitz_handbook_1972}, and $\alpha>0$ is introduced as a scaling
parameter. The Sturmian functions $\{S_n^{(\alpha)}(r),n\geq1\}$ form a
complete orthonormal set with respect to the weight function $1/r$ on the
interval $I=[0,+\infty[$ (for a detailed discussion of completeness properties,
see \eg Ref.~\cite{weniger_weakly_1985}). In particular, it is worth noting that
$S_N^{(N/Z)}(r)$ coincides up to a normalization factor with the wave function
of the $N$th 1D hydrogenic bound state in the Coulomb potential $-Z/r$. This
implies that for a specific scaling $\alpha$ one such hydrogenic state may be
represented faithfully by a single element of the Sturmian basis. While the
complete (yet discrete) Sturmian basis is able to represent all bound and
continuum states of the hydrogen or helium atom, in any practical computation
the basis must be truncated to contain only functions below a certain
$N_{\mathrm{max}}$. This truncation introduces a soft finite-size cut-off at a
typical distance of $2 \alpha N_{\mathrm{max}}$ Bohr radii. This distance
corresponds to the approximate position of the outermost extremum of
$S_{N_{\mathrm{max}}}^{(\alpha)}(r)$, as given by
expression~(\ref{eq:1Dsturmians}). Because of this cut-off, the atomic continua
are discretized into quasi-continua, Rydberg series are resolved only up to a
certain degree of excitation, and the calculated ionization thresholds are
shifted slightly downwards with respect to the exact values (see
Refs.~\cite{krug_alkali_2001, schlagheck_drei-korper-coulombproblem_1999} and
below). By increasing $N_{\mathrm{max}}$ and adapting $\alpha$, however, these
effects can be made small.

% Finally, note that the
% expansion~(\ref{eq:phitheta_expansion}), the behavior of
% $S_n^{(\alpha)}(r)$ close to the origin, and the
% ansatz~(\ref{eq:phi_theta}) imply certain boundary conditions for the
% eigenvalue problem~(\ref{eq:evp_perimetric}), inasmuch as
% $\phi_{>}^\theta(x,y)$ vanishes linearly in $x$ or $y$ when $x$ or $y$
% is sent to zero keeping the other distance fixed, and
% $\phi_{>}^\theta(x,y)=xy(x+y)(C+o(x,y))$, with $C$ constant, when both
% $x$ and $y$ become small. While the exact asymptotic behavior of the
% two-electron wave function in the close vicinity of the
% triple-collision point remains a complicated matter (see \eg
% Ref.~\cite{popov_rigorous_2000, byun_scaling_2007} and references
% therein for the 3D case), the combination of
% (\ref{eq:generalized_evp}) and (\ref{eq:phitheta_expansion}) has been
% found to produce very accurate results for the spectrum of \Zee{}
% helium~\cite{schlagheck_drei-korper-coulombproblem_1999}.

Using a Sturmian basis set has the following advantage from the computational
point of view: The operators
%%%%
\begin{eqnarray}
S_{\pm}^{(\alpha)}&=
\frac{r}{2\alpha}
+\frac{\alpha r}{2}\frac{\diffd^2}{\diffd r^2}
\mp r\frac{\diffd}{\diffd r}
\label{eq:Spm_action}\\
S_{3}^{(\alpha)}&=
\frac{r}{2\alpha}
-\frac{\alpha r}{2}\frac{\diffd^2}{\diffd r^2}
\label{eq:S3_action}
\end{eqnarray}
%%%%
form a Lie algebra of ladder operators with the commutators
$[S_{-}^{(\alpha)},S_{+}^{(\alpha)}]=2S_3^{(\alpha)}$ and
$[S_3^{(\alpha)},S_\pm^{(\alpha)}]=\pm S_\pm^{(\alpha)}$, and the following
action on the Sturmian functions $S_n^{(\alpha)}(r)=\braket{r}{S_n^{(\alpha)}}$
of equation~(\ref{eq:1Dsturmians})~\cite{delande_group_1984,
delande_atomes_1988}:
%%%%
\begin{eqnarray}
S_{\pm}^{(\alpha)} \ket{S_n^{(\alpha)}}&=
\sqrt{n(n\pm1)}\ket{S_{n\pm1}^{(\alpha)}}\\
S_{3}^{(\alpha)} \ket{S_n^{(\alpha)}}&=
n\ket{S_n^{(\alpha)}}.
\end{eqnarray}
%%%%
To exploit these properties we use product bases of two sets of
Sturmian functions that are fitted to the specific configuration.

For the \eZe{} configuration we expand $\psi^{\theta}(z_1, z_2)$ on a
(anti-) symmetrized product basis as
\begin{eqnarray}
    \label{eq:phitheta_eZe_expansion}\fl
    \psi^{\theta}_{\mathrm{odd}}(z_1, z_2)
    &= \sum_{n_1=1}^{N}\sum_{n_2=n_1+1}^N
    \frac{C_{n_1n_2}}{\sqrt{2}}
    (S_{n_1}^{(\alpha)}(z_1)S_{n_2}^{(\alpha)}(z_2)
    - S_{n_1}^{(\alpha)}(z_2)S_{n_2}^{(\alpha)}(z_1)),\\\fl
    \psi^{\theta}_{\mathrm{even}}(z_1, z_2)
    &= \sum_{n_1=1}^{N}\sum_{n_2=n_1+1}^N
    \frac{C_{n_1n_2}}{\sqrt{2}} (S_{n_1}^{(\alpha)}(z_1)S_{n_2}^{(\alpha)}(z_2)
    + S_{n_1}^{(\alpha)}(z_2)S_{n_2}^{(\alpha)}(z_1))\\\fl
    &+ \sum_{n_1=1}^{N}
    C_{n_1n_1} S_{n_1}^{(\alpha)}(z_1)S_{n_1}^{(\alpha)}(z_2),
\end{eqnarray}
for the odd and even states respectively.
It is now our goal to express the operators in the eigenvalue equation
as polynomials in the ladder operators. This is not immediately
possible, because they admit no representation of $\frac{1}{r}$. To
overcome this problem we multiply the eigenvalue equation from the
left by a factor of $e^{3\ci\theta}z_1z_2(z_1 + z_2)$. Note that for
\eZe{} we have $|z_1 \mp z_2| = (z_1 + z_2)$. This leaves us with the
generalized eigenvalue problem
\begin{equation}
    \label{eq:generalized_evp_eZe}
    A_{\theta}\psi^{\theta}(z_1, z_2) =
    E^{\theta}B_{\theta}\psi^{\theta}(z_1, z_2),
\end{equation}
with
%%%%
\begin{eqnarray}
 A_\theta&=e^{3\ci\theta}z_1 z_2 (z_1 + z_2) H_\theta
 \label{eq:generalized_evp_opA_eZe}\\
 B_\theta&=e^{3\ci\theta}z_1 z_2 (z_1 + z_2)\label{eq:generalized_evp_opB_eZe}.
\end{eqnarray}
%%%%
Consequently all the fractions appearing in the Hamiltonian cancel. Then, with the help of  expressions~(\ref{eq:Spm_action}) and~(\ref{eq:S3_action}), the operators $A_\theta$ and $B_\theta$ in equations~(\ref{eq:generalized_evp_opA_eZe}) and~(\ref{eq:generalized_evp_opB_eZe}) are written as polynomials of degree three in $S_\pm^{(\alpha_i)}$ and $S_3^{(\alpha_i)}$. The small degree of these polynomials enforces strong selection rules between the elements of the product basis. As a result, the matrix representations of the operators of equation (\ref{eq:generalized_evp_eZe}) in the Sturmian basis are very sparse, and the eigenvalue problem~(\ref{eq:generalized_evp_eZe}) is amenable to standard numerical techniques based on Krylov subspaces~\cite{hernandez_slepc:_2005}.

In \Zee{} helium we exploit the stronger symmetry that was discussed
in Section~\ref{sec:Zee_wavefunctions} in a different way.
Owing to the exchange symmetry and the singularity of the Coulomb potential
between electrons, the two-electron \Zee{} wave function separates into
%%%%
\begin{equation}
    \label{eq:separating-wave-function}
    \psi(z_1,z_2)=\psi_{>}(z_1,z_2)\pm\psi_{>}(z_2,z_1),
\end{equation}
%%%%
where $\psi_{>}(z_1,z_2)$ is defined on the domain $0<z_2<z_1$, and vanishes at
its boundaries. On this domain, upon introduction of the {\em perimetric
coordinates}
%%%%
\begin{eqnarray}
x&=z_1-z_2\\
y&=z_2,
\end{eqnarray}
%%%%
one is left with the eigenvalue problem
%%%%
\begin{equation} \label{eq:evp_perimetric}
H_\theta \phi_{>}^\theta(x,y)=E^\theta\phi_{>}^\theta(x,y),
\end{equation}
%%%%
with $\phi_{>}^\theta(x,y)=\psi_{>}^\theta(x+y,y)$ and
%%%%
\begin{equation} \label{eq:Hxytheta}\fl
    H_\theta = e^{-2\ci\theta}\left(
    -\frac{\partial^2}{\partial x^2}
    -\frac{1}{2}\frac{\partial^2}{\partial y^2}
    +\frac{\partial^2}{\partial x\partial y}
    \right)
    +e^{-\ci\theta}\left(
    -\frac{Z}{x+y}-\frac{Z}{y}+\frac{1}{x}
    \right).
\end{equation}
%%%%
Following Ref.~\cite{schlagheck_drei-korper-coulombproblem_1999},
equation~(\ref{eq:evp_perimetric}) is replaced by the generalized eigenvalue
problem
\begin{equation}
    \label{eq:generalized_evp}
    A_{\theta}\phi^{\theta}(x, y) =
    E^{\theta}B_{\theta}\phi^{\theta}(x, y),
\end{equation}
where
\begin{eqnarray}
    A_\theta&=e^{4\ci\theta} x y (x+y) H_\theta
    (x+y)\label{eq:generalized_evp_opA}\\
    B_\theta&= e^{4\ci\theta} x y
    (x+y)^2\label{eq:generalized_evp_opB},
\end{eqnarray}
%%%%
which is obtained upon multiplication of equation (\ref{eq:evp_perimetric}) by
$e^{3\ci\theta}xy(x+y)$ from the left, with the substitution
%%%%
\begin{equation} \label{eq:phi_theta}
\phi_{>}^\theta(x,y)=e^{\ci\theta}(x+y)\phi^\theta(x,y).
\end{equation}
%%%%
Now $\phi^\theta(x,y)$ is expanded on a
Sturmian product basis as
%%%%
\begin{equation} \label{eq:phitheta_expansion}
\phi^\theta(x,y)=
\sum_{n_x=1}^{N_x} \sum_{n_y=1}^{N_y}
C_{n_x n_y} S_{n_x}^{(\alpha_x)}(x) S_{n_y}^{(\alpha_y)}(y),
\end{equation}
%%%%
where $N_x$ and $N_y$ arise from the truncation of the basis in any practical
computation, and $\alpha_x$ and $\alpha_y$ are scaling factors of the basis in
the $x$ and $y$ directions. The latter can be adapted independently for an
optimal representation of certain electronic configurations in the truncated
basis.
In this respect, note that
multiplying $H_\theta$ from the left by $e^{3\ci\theta} x y (x+y)$,
without additionnal $e^{\ci\theta}(x+y)$ factor from the right,
actually suffices for a polynomial representation of the resulting
operator in terms of the ladder operators $S_\pm^{(\alpha_i)}$ and
$S_3^{(\alpha_i)}$. In that case, the total degree of the polynomial
is only three and the selection rules are even stronger (the operators
$A_\theta^\prime=e^{3\ci\theta} x y (x+y) H_\theta$ and
$B_\theta^\prime=e^{3\ci\theta} x y (x+y)$ of the resulting eigenvalue
problem couple each element of the Sturmian basis to a maximum of 20
other elements, with $|\Delta n_x|\leq 2$ and $|\Delta n_y|\leq
2$).
Note however, that the matrix representation of $A_\theta^{\prime}$ in the
Sturmian basis is \textit{not} complex-symmetric, which limits the
efficiency of suitable eigenvalue routines.

In this way we have described both 1D configurations as sparse eigenvalue
problems that are accessible for standard numerical techniques. For the
calculation of the eigenvalues and eigenvectors, we used the SLEPc
library~\cite{hernandez_slepc:_2005}, which itself is build on top of
PETSc~\cite{balay_petsc_2013-1, balay_efficient_1997}, in combination
with the parallel solver MUMPS~\cite{amestoy_fully_2001,
  amestoy_hybrid_2006}.

%%%%
\begin{figure*}
    \includegraphics[width=\textwidth]{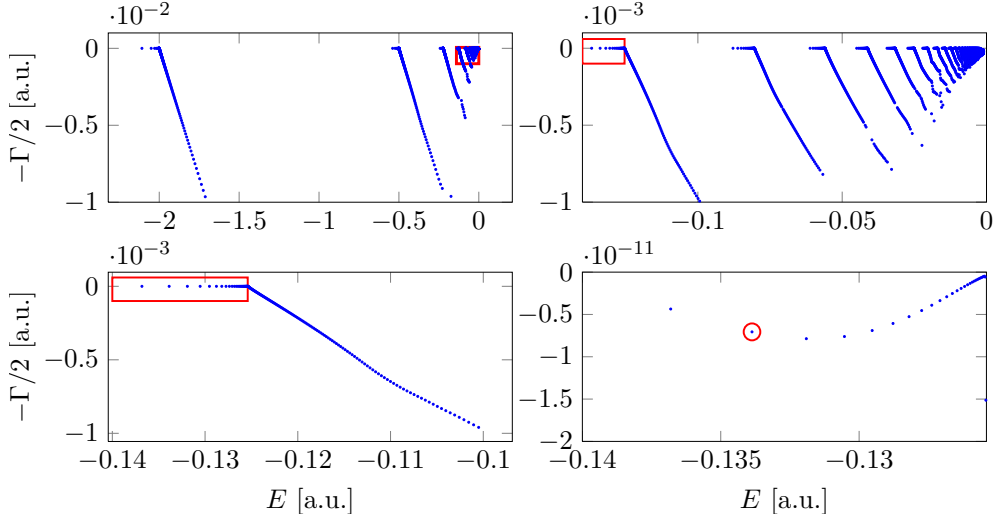}%
    \caption{\label{fig:series_4_spectrum}
      (Colour online) Spectrum of 1D helium in the \Zee{}
      configuration under complex rotation. The eigenvalues in blue
      are obtained from the solution of the eigenvalue
      problem~(\ref{eq:generalized_evp}) in the Sturmian basis through
      the Krylov-Schur method, with numerical parameters
      $\theta=0.005$, $\alpha_x=50$, $\alpha_y=50$, $N_x=300$,
      $N_y=150$. The top left panel gives an overview of the spectrum
      with the ground state at -2.108~\au and the double-ionization
      threshold (DIT) at zero energy. The following panels (from left
      to right, and top to bottom) are consecutive zooms into the
      marked areas. The different series corresponding to states of
      the inner electron are visible with their rotated (and
      discretized) continua (see
      sections~\ref{subsec:complex_rotation} and
      \ref{subsec:numerics}). The bound states below the first
      single-ionization threshold at
      $E_1^\mathrm{\SIT}=-2~\mathrm{\au}$ lie on the real axis,
      whereas the resonances above $E_1^\mathrm{\SIT}$ have a small
      imaginary part, as shown by the bottom right panel. The two last
      panels (bottom) show the $N=4$ series, i.e. the series
      corresponding to the third excited state of the inner
      electron. The bottom right graph highlights the (4,6) resonance
      with a red circle. This resonance has the eigenvalue
      $E_{(4,6)}=(-1.3387\cdot10^{-1}-7.06\cdot10^{-12}
      \ci)~\mathrm{\au}$, i.e. an energy of $-0.13387~\mathrm{\au}$
      and a decay rate of $1.41 \cdot 10^{-11}~\mathrm{\au}$}.
\end{figure*}
%%%%

\begin{figure}
    \includegraphics[width=0.5\textwidth]{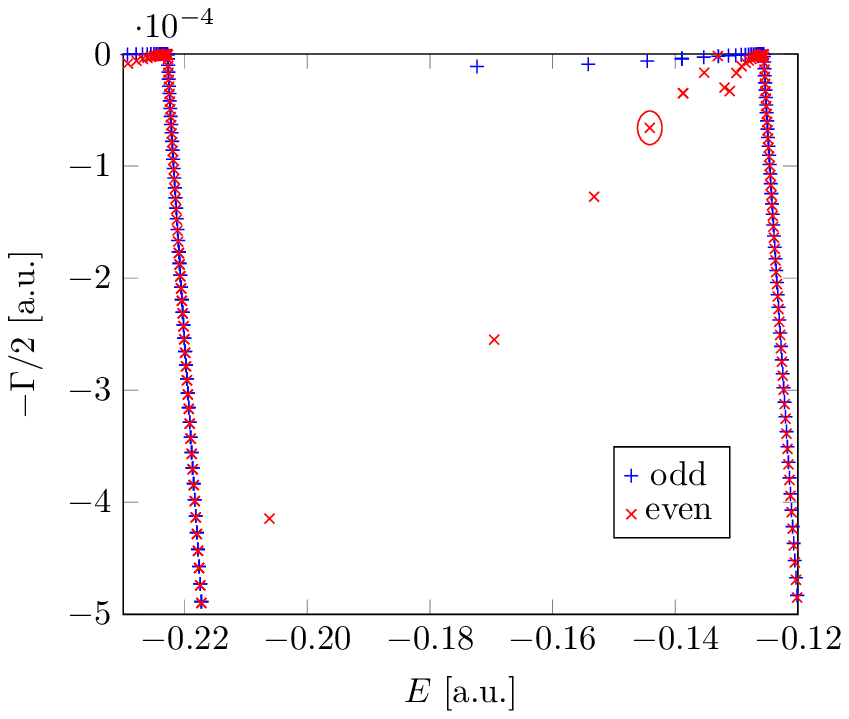}
    \caption{\label{fig:series_4_eZe_spectrum}
      (Colour online) Spectrum of 1D helium in the \eZe{}
      configuration under complex rotation. This part of the spectrum
      shows the $N=4$ series of resonances between the $N=3$ continuum
      on the left and the $N=4$ continuum on the right. The blue + symbols
      refer to the odd states,
      the red x symbols to those with even symmetry (see discussion
      in section \ref{sec:Zee_wavefunctions}). While there is almost a one-to-one 
      correspondence between the two symmetries for the
      resonances, the even states exhibit much higher decay
      rates. Furthermore, since the (4,4) state is only possible in
      the even symmetry, this series has one more state. This is also
      the reason for the deviation from the smooth behaviour of the
      resonances approaching the threshold:
      Around $E=-0.13~\mathrm{\au}$ we see the intrusion
      of a perturber from the $N=5$ series in the even
      configuration. In the odd symmetry this only happens in even
      higher series. The red circle marks the (4,7,e) resonance. It
      will be analysed in more detail below, and has an
      eigenvalue of $E_{(4,7,e)} = (-1.4415 \cdot 10^{-1} - 6.5915
      \cdot 10^{-5} \ci)~\mathrm{\au}$, i.e. an energy of
      $-0.14415~\mathrm{\au}$ and a decay rate of $1.3183 \cdot
      10^{-4}~\mathrm{\au}$
    }
\end{figure}

Figure~\ref{fig:series_4_spectrum} shows a complex-rotated \Zee{} spectrum
obtained via the Krylov-Schur method as described in
Ref.~\cite{hernandez_krylov-schur_2007}. This spectrum displays the features
discussed in section~\ref{subsec:complex_rotation}, in particular the isolated
eigenvalues $E_i^\theta=\Re (E_i^\theta)-\ci\Gamma_i/2$ in the lower half of the
complex plane ($\Gamma_i>0$), which correspond to resonance states (see bottom
right panel).
Figure~\ref{fig:series_4_eZe_spectrum} shows a similar spectrum for
the \eZe{} case. The two series associated with even and odd symmetry,
respectively, can be clearly distinguished.

For the further analysis with regards to partial rates, we need to
calculate wave functions from the now available eigenvectors. To this
end we exploit the second essential advantage brought along by the
Sturmian basis set, namely that the matrix elements of the
complex-rotation operator~(\ref{eq:Rtheta}) between Sturmians are
known analytically~\cite{englefield_group_1972, delande_group_1984,
  delande_atomes_1988}:
%%%%
\begin{eqnarray} \label{eq:matel_backrotation}
\matel{\widetilde{S}_n^{(\alpha)}}{R(-\theta)}{S_{n'}^{(\alpha)}}
&=&e^{\ci\theta/2}\ci^{n-n'}\sqrt{n n'}
\left(\cos\frac{\theta}{2}\right)^{-(n+n')}
\left(\sin\frac{\theta}{2}\right)^{n+n'-2}\nonumber\\
&&\quad\times
{}_2F_1\left[-n+1,-n'+1;2;\left(\sin\frac{\theta}{2}\right)^{-2}\right],
\end{eqnarray}
%%%%
where we use the notation
$\braket{\widetilde{\phi}}{\psi}=\int_0^{+\infty}r^{-1}\overline{\phi(r)}\psi(r)\diffd
r$, and ${}_2F_1$ is the hypergeometric
function~\cite{abramowitz_handbook_1972}.

\begin{figure*}
    \includegraphics[width=\textwidth]{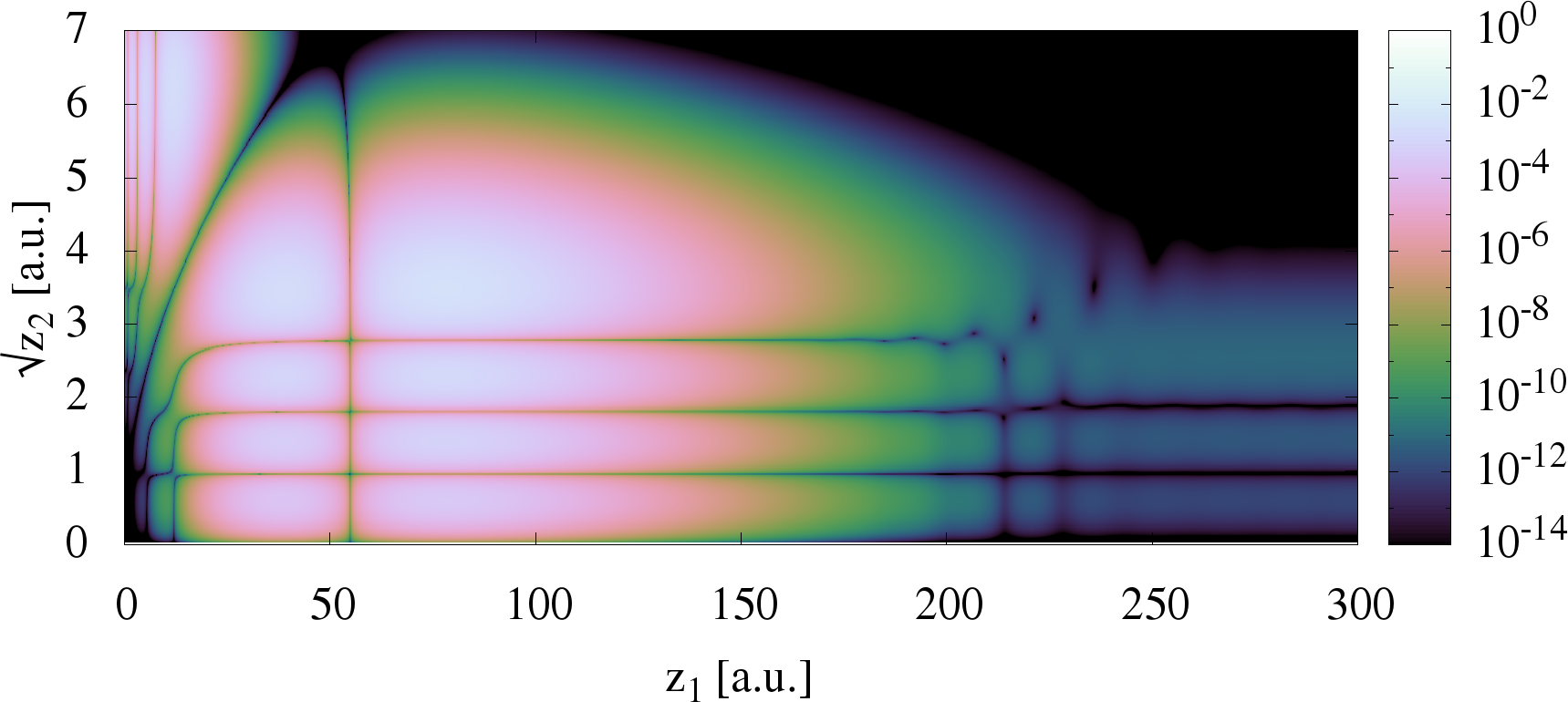}
    \caption{\label{fig:resonance_4_6}
      Electronic density $|\psi(z_1,z_2)|^2$ of the \Zee{} $(4,6)$ resonance,
      as obtained upon back-rotation of the associated resonance wave
      function $\ket{\phi_{>}^\theta}$ (see
      section~\ref{subsec:numerics}) and extension to the entire
      $(z_1,z_2)$ domain. The numerical parameters are $\theta=0.001$,
      $\alpha_x=2$, $\alpha_y=2$, $N_x=6000$, $N_y=300$.
      The $z_2$ direction is scaled with the square root to improve
      the visibility of the nodal structure. The clearly visible nodal
      line on the $z_1=z_2$ diagonal is distorted accordingly. The density
      profile reveals a central part, that quickly tails off
      exponentially after its maximum at around $z_1 =
      100~\mathrm{\au}$,
      which corresponds to a bound state of the two electrons around
      the nucleus, and an asymptotic part along the $z_1$ axis
      ($z_1\gtrsim250~\mathrm{\au}$) which indicates the admixture of
      singly ionized states.}
\end{figure*}

The exact calculation again differs for the two configurations. For the
\Zee{} case the representation of $R(-\theta)\ket{\phi_i^\theta}$ in the
Sturmian basis gives the back-rotated resonance wave function
$\phi_{>,i}(x,y) = \matel{x,y}{R(-\theta)}{\phi_{>,i}^\theta} =
\matel{x,y}{R(-\theta)e^{\ci\theta}(x+y)}{\phi_{i}^\theta} =
(x+y)\matel{x,y}{R(-\theta)}{\phi_{i}^\theta}$ on the perimetric
domain without analytic continuation, by evaluating the Sturmian functions with
real arguments $x$ and $y$. The resulting spatial wave function
$\psi_{>,i}(z_1,z_2)=\phi_{>,i}(x,y)$ is extended to the full \Zee{} domain by
simple symmetrization, which yields $\psi_i(z_1,z_2)$.
Figure~\ref{fig:resonance_4_6} shows the electronic density $|\psi(z_1,z_2)|^2$
obtained upon back-rotation of the $(4,6)$ resonance state (marked by a red
circle in Fig.~\ref{fig:series_4_spectrum}) on the entire $(z_1,z_2)$ domain.
%%%%
%%%%
The density plot (on a logarithmic scale) has two important features: a central
part that extends up to $100-200~\mathrm{\au}$ in $z_1$ direction and
an asymptotic
contribution that runs along the $z_1$ axis beyond $250~\mathrm{\au}$ The
central part corresponds to the remnant of a bound state of the two electrons
around the nucleus. Counting the nodal lines along both directions confirms the
approximate quantum numbers. At distance $z_1=100~\mathrm{\au}$, for instance,
Fig.~\ref{fig:resonance_4_6} reveals three nodes in $z_2$ direction, in
agreement with $N=4$. A closer inspection of the region around the origin (not
shown here) also confirms that the central part displays four nodes in
$z_1$ direction, close to the $z_1$ axis. The asymptotic part, on the
other hand, corresponds to the unbound motion of one of the electrons, and
signals the admixture of singly ionized states. Interestingly, the asymptotic
density profile exhibits a clear nodal structure along the $z_2$ direction, as
in the bound part. In the transition regime between $200~\mathrm{\au}$ and
$250~\mathrm{\au}$ one nodal line vanishes. This indicates that in the final
state of the autoionization process, the remaining electron has dropped to level
three of the helium ion $\textrm{He}^+$.

\begin{figure*}
    \includegraphics[width=\textwidth]{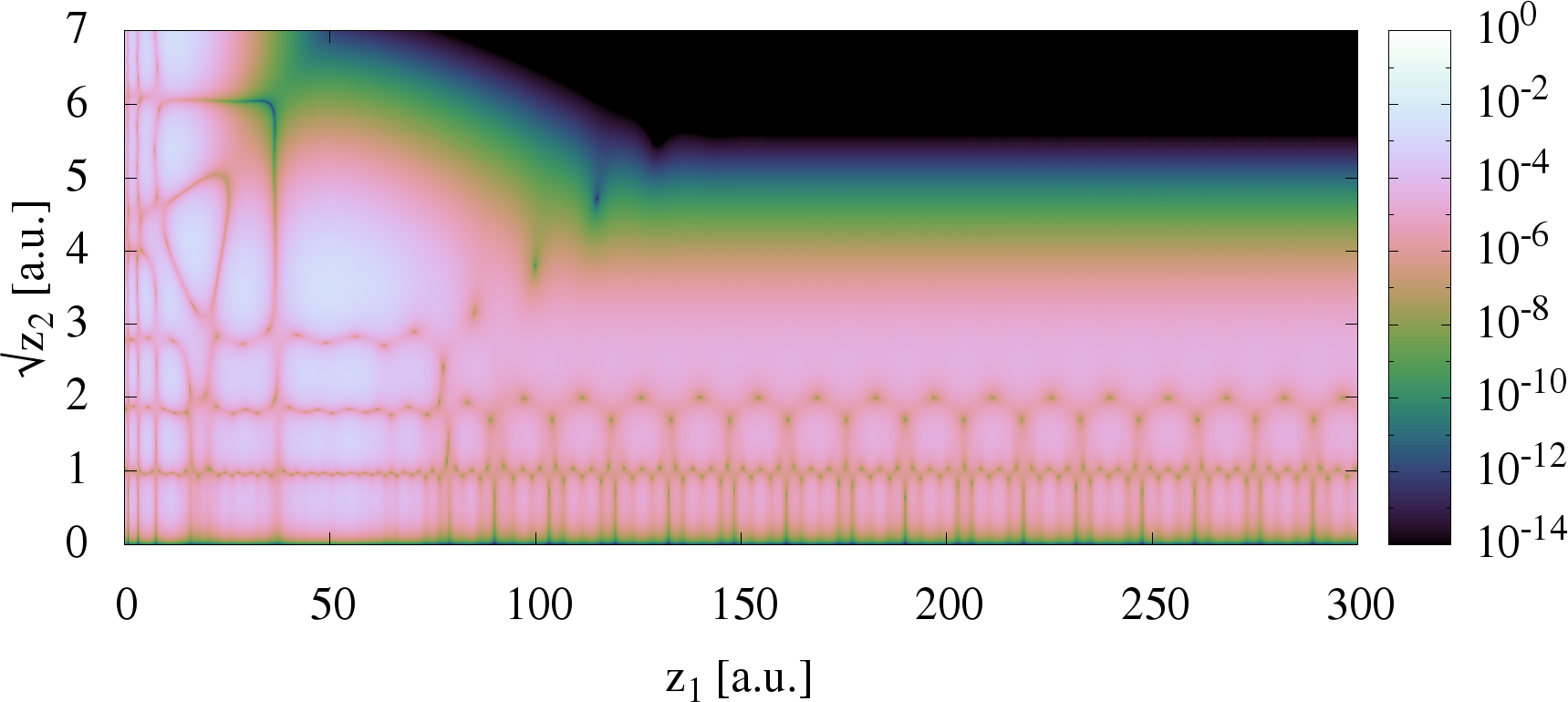}
    \caption{\label{fig:resonance_4_7_e}
      Electronic density $|\psi(z_1,z_2)|^2$ of the \eZe{} $(4,7,e)$
      resonance, as obtained upon back-rotation of the associated
      resonance wave function $\ket{\phi_{>}^\theta}$ (see
      section~\ref{subsec:numerics}). The numerical parameters are
      $\theta=0.005$, $\alpha=2$, $N=1500$. The $z_2$ direction is
      scaled with the square root to improve the visibility of the
      nodal structure. The density profile reveals a central part
      ($z_1 \lesssim 100~\mathrm{\au}$) which corresponds to a bound
      state of the two electrons around the nucleus, and an asymptotic
      part along the $z_1$ axis ($z_1 \gtrsim 150~\mathrm{\au}$) which
      indicates the admixture of singly ionized states.}
\end{figure*}

For the \eZe{} case we start by transforming the coefficients in the
(anti-) symmetrized basis into the ordinary product basis of Sturmians
for $z_1$ and $z_2$. The evaluation of the spatial wave function is
then done as in the \Zee{} case, without the additional $(x+y)$
factor and immediately on the whole domain. The resulting electronic
density $|\psi(z_1,z_2)|^2$ for the $(4,7,\mathrm{even})$ state is
depicted in Fig.~\ref{fig:resonance_4_7_e}.
%%%%
%%%%
It shares many features with its \Zee{} counterpart. There are,
however, several differences: Since the state is even, there is no
nodal line along the diagonal. Instead here we find an antinode, only
interrupted by the nodes in $z_1$ or $z_2$ direction. The density also
shows a central and an asymptotic part. The extension of the central
part is considerably smaller, which is unsurprising, since the two
electrons here are on opposite sides of the nucleus. As compared to
Fig.~\ref{fig:resonance_4_6}, the asymptotic part, now starting around
$100~\mathrm{\au}$, has a much larger weight. This is a result of the
far larger decay rates of the \eZe{} configuration. Additionally there
are further variations or ripples in the asymptotic part. These are an
effect of the presence of several open channels and will be examined
more closely in Sect.~\ref{sec:partial_rates_Zee}.

The above examples demonstrate that in contrast to the complex-rotated
spectrum, which only gives access to the total decay rate of
individual resonance states, an inspection of the back-rotated
resonance wave function may reveal how those resonance states
decay. While the presented resonance wave functions lend themselves to
a clear qualitative interpretation, the analysis of decaying resonance
states can be put on a systematic and quantitative basis by defining
channel wave functions and computing the associated partial decay
rates. In the following section, we show how the computation of such
partial decay rates can be achieved in the setting of complex rotation.

\section{Partial decay rates from complex
  dilation\label{sec:partial_rates}}

In this section we present a general method for the calculation of
partial decay rates. First we recall the calculation of rates from
configuration-space analysis of wave functions. Then we demonstrate
how wave functions can be decomposed into partial wave functions in a
way compatible with the calculation of rates. This allows for the
calculation of partial rates.

\subsection{Total decay rates and flux densities\label{subsec:fluxes}}
A decay rate is the rate at which the probability of a state in
some region of configuration space decreases. More specifically, in
this case we are looking at bound continuum transitions. Therefore the
rate we are interested in is the rate of loss of probability from the
region of configuration space that is associated with an initial
state, here a bound state.

First we describe the calculation of rates from wave functions in
configuration space. The setup is as follows: We have a configuration
space $C$, whose coordinates we write as $\vecop{r}$. Furthermore we have a
time dependent state $\ket{\psi_t}$. Due to local conservation of
probability density (see \eg
\cite{cohen-tannoudji_quantenmechanik_1997}), we have the continuity
equation
\begin{equation}
    \label{eq:continuity-equation}
    \frac{\partial}{\partial t}|\psi_t(\vecop{r})|^2 = -\nabla
    \vecop{j}_t(\vecop{r}),
\end{equation}
where $\vecop{j}$ is the probability current density.

We are interested in the decay of the state. Therefore we assume the
existence of a reaction region, for example the spatial region where the
electronic density is localized at the onset of an ionization process. Due to the
long-range nature of the Coulomb potential, this region is here not bounded. We
choose a compact volume $V_R \subset C$ that is centered on this
reaction region and parameterized by a length scale $R$, such that
$\lim_{R\to\infty}V_R = C$. The intuitive picture is the sphere of
appropriate dimension and radius $R$. However we will see that other
shapes can be advantageous.

By integration over $V_R$ and application of Gauss' theorem, the loss of probability density inside $V_R$ is given by
\begin{equation}
    \label{eq:continuity-integrated}
    \int_{V_R} \frac{\diffd}{\diffd t} |\psi_t|^2\diffd V =
    -\oint_{\partial V_R}
    \vecop{j}_t \diffd\vecop{S}.
\end{equation}
If the state exhibits a simple exponential decay with decay rate
$\Gamma$ we know that
\begin{equation}
    \label{eq:exponential-decay}
    \frac{\diffd}{\diffd t} |\psi_t|^2 = -\Gamma |\psi_t|^2,
\end{equation}
hence if we evaluate (\ref{eq:continuity-integrated}) for such
a state we find
\begin{equation}
    \label{eq:local-rate}
    \Gamma = \frac{\oint_{\partial V_R} \vecop{j}_t \diffd\vecop{S}}
{\int_{V_R} |\psi_t|^2\diffd V},
\end{equation}
\ie the ratio on the right hand side is constant and independent of
$R$, even though both numerator and denominator might diverge for
$R\to\infty$. The denominator, \ie the probability density inside
$V_R$, is readily integrated for a given $R$. Thus the remaining task is the
calculation of flux of the probability current density $\vecop{j}$
through $\partial V_R$. Note that it is sufficient to calculate the
component of $\vecop{j}$ that is perpendicular to the
boundary. To account for this we define $\tilde{j} :=
\vecop{j}\vecop{n}$, where $\vecop{n}$ is the surface normal, \ie
$\diffd\vecop{S} = \vecop{n}\diffd S$.

For a standard discussion of current densities in the quantum-mechanical setting
we refer to the literature (\eg \cite{cohen-tannoudji_quantummechanics1_2005}).
Here we only recall the necessary results: The current density
$\vecop{j}_t(\vecop{r})$ at point $\vecop{r}$ and time $t$ is obtained as the
expectation value of the current-density operator $\vecop{J}(\vecop{r})$, \ie
\begin{eqnarray}
    \label{eq:flux-density-operator}
    \vecop{j}_t(\vecop{r}) =
    \bra{\psi_t}\vecop{J}(\vecop{r})\ket{\psi_t}, \\
    \vecop{J}(\vecop{r}) =
    \frac{1}{2}\left(\vecop{p}\ket{\vecop{r}}\bra{\vecop{r}} +
    \ket{\vecop{r}}\bra{\vecop{r}}\vecop{p}\right).
\end{eqnarray}
This naturally leads to the current-density operator in direction of
the boundary surface normal
\begin{equation}
    \label{eq:projected-density-current-operator}
    \tilde{J}(\vecop{r}) =
    \frac{1}{2}\left(\tilde{p}\ket{\vecop{r}}\bra{\vecop{r}} +
    \ket{\vecop{r}}\bra{\vecop{r}}\tilde{p}\right).
\end{equation}

In the context of back-rotated wave functions from complex rotation,
$\ket{\psi}=R(-\theta)\ket{\psi_{i_0}^\theta}$ is not normalizable. Yet, the
associated probability density
$\braket{\vecop{r}}{\psi}\braket{\psi}{\vecop{r}}=|\psi(\vecop{r})|^2$ is finite
everywhere. Furthermore, as discussed
in section~\ref{subsec:complex_rotation}, the back-rotated wave function is
still a solution to the time-independent Schr\"odinger equation with eigenvalue
given by the complex eigenvalue of the rotated wave function. With the usual
time evolution it therefore also generates a solution of the time-dependent
Schr\"odinger equation with a purely exponential decay. Hence the normal
derivation of the current-density operator works as well as the calculation of
the current density as its expectation value.
%The only difference that should be noted for the developments below is the
%substitution of $\bra{\psi}$ in the case of a physical state by
%$\bra{\overline{\psi}}$ in the case of back-rotated resonance wave functions.
%For simplicity, we shall keep the notation
%$\bra{\psi}$ for both cases.

Thus the derivation is complete: Given a wave function in
configuration space that exhibits a purely exponential decay we can
calculate the decay rate $\Gamma$ by analyzing the probability current
density as given by the expectation value of the current density
operator according to (\ref{eq:local-rate}).
Since the decay rate is here assumed constant in time, we now drop the
index $t$ and always imply $t=0$.
Let us now decompose the wave function to access partial rates.

\subsection{Partial rates\label{subsec:partial_rates}}

Partial rates are rates associated with particular decay
channels. What constitutes a channel can vary with the problem at
hand, but will usually be defined by some condition on a part of the
system. It is thus nothing but a subspace in Hilbert space. We treat
the various channels by the projectors onto these subspaces.

To show that this perspective leads to a useful concept of partial
rates, let us assume that we have a family of projectors $P_N$, such
that
\begin{eqnarray}
    \label{eq:channel-completeness}
    \sum_NP_N &= \bbone\\
    P_NP_M &= \delta_{NM}P_N,
\end{eqnarray}
\ie the corresponding subspaces are orthogonal and span the Hilbert
space.

The goal is now to split the right hand side of
(\ref{eq:local-rate}) into a sum of terms that can be
connected with the decay into given channels. For this we keep the
denominator and split the numerator according to the channels.
\begin{eqnarray}
    \fl
    \oint_{\partial V_R} \vecop{j}(\vecop{r})
    \diffd\vecop{S}(\vecop{r})
    &= \oint_{\partial V_R} \matel{\psi}{\tilde{J}(\vecop{r})}{\psi}
    \diffd S\\
    &= \sum_{N} \oint_{\partial V_R}
    \matel{\psi}{P_N^2\tilde{J}(\vecop{r})}{\psi} \diffd S\\
    \label{eq:partial-rates}
    &= \sum_{N} \oint_{\partial V_R}
    \matel{\psi}{P_N\tilde{J}(\vecop{r})P_N}{\psi} \diffd S
    - \sum_{N} \oint_{\partial V_R}
    \matel{\psi}{P_N[\tilde{J}(\vecop{r}),
      P_N]}{\psi} \diffd S.
\end{eqnarray}
The second term of this sum will vanish if we choose $V_R$ compatible
with the channels. We will see what this means more clearly in the
application of the method in section~\ref{sec:partial_rates_Zee}.

We introduce the projected wave function $\ket{\psi_N} :=
P_N\ket{\psi}$ and define partial rates $\Gamma_N(R)$ as
\begin{equation}
    \label{eq:definition-partial-rates}
    \Gamma_N(R) = \frac{\oint_{\partial V_R}
      \matel{\psi_N}{\tilde{J}(\vecop{r})}{\psi_N} \diffd S}
    {\int_{V_R} |\psi|^2\diffd V}.
\end{equation}
This is not a straightforward application of
(\ref{eq:local-rate}) to the projected wave functions, since
the denominator is in fact calculated with the full wave
function. Still, if $V_R$ is chosen such that the second term in
(\ref{eq:partial-rates}) vanishes, we see that the partial rates
sum up to the total rate:
\begin{equation}
    \label{eq:local-partial-rates}
    \Gamma = \sum_{N} \Gamma_N(R).
\end{equation}

Note that these partial rates depend on the extension of $V_R$. Their
sum, however, the total rate $\Gamma$, stays constant
for all $R$. Furthermore this gives us a well defined way to evaluate
the partial rates for any $R$, and thus enables access to the
asymptotic behaviour, without any ad-hoc assumptions on the final
states. In other words, we can calculate the involved quantities as
functions of $R$ and identify the asymptotic region as the part where
no more interplay among the channels is observed, directly from the
data. This is in contrast to other approaches that assume a specific
final state both for the expelled particle and the remaining system
\cite{goldzak_evaluation_2010, moiseyev_partial_1990,
  scrinzi_ionization_1997, foumouo_theory_2006,
  nikolopoulos_time-dependent_2007, argenti_photoionization_2013,
  menzel_decay_1995}.
Since we want to study these quantities as functions of $R$, the
following definitions will be useful:
\begin{eqnarray}
    \label{eq:definition-local-quantities}
    \gamma_{(N)}(R) = \oint_{\partial V_R}
    \matel{\psi_{(N)}}{\tilde{J}(\vecop{r})}{\psi_{(N)}} \diffd S\\
    D(R) = \int_{V_R} |\psi|^2\diffd V.
\end{eqnarray}

It is at this point useful to recall that the back-rotated
wave function of a resonance computed with complex rotation is a
static representation of a dynamic process. Starting from an initial
state that corresponds to a bound state, one or several decay
processes take place that lead to a net outgoing flux of probability
density. This, in turn, generates a spatially unbounded density
distribution. In the case of an expelled particle, this gives the
probability to find the particle in some (infinitesimal) space
region. We have up to now neglected the time domain. To reconcile the
two pictures, we recognize that time and space in the case of a moving
classical particle with fixed energy are linked rather directly as
$\Delta_x = v \Delta_t$, where $v$ is the velocity of the particle.
In this sense a point in space outside the bound part of
the wave function can be associated with a time interval that must
have passed since the expulsion of the particle.
The full quantum picture is more complicated due to the uncertainty of
the momentum and the dispersion of the associated wave packet. The
general picture, however, still holds, see the discussion in
\cite[Sect.~2.1]{ho_method_1983}.

The continuing decay
of the wave function according to (\ref{eq:exponential-decay}) means
that the reference density of the bound part of the system decreases
exponentially in time. Consequently the resonance wave function in its
asymptotic part must grow exponentially in space, since more distant
points in space are linked to earlier times and thus to larger initial
densities.

This poses no problem in the case of a single channel, since the
normalization with the integrated density accounts for exactly this
effect. This is the physical reason why the ratio in
(\ref{eq:local-rate}) is well behaved. In the case of several open
channels we face a challenge: Since the total energy, that is of
course constant and dictated by the real part of the resonance, is
distributed differently to system and expelled particle for the
different channels, the escaping particles have different velocities
and therefore the partial currents grow exponentially with different
rates. In this sense points in the asymptotic region mix contributions
from the different channels at different times. Since the density
normalization is such that the total rate stays constant, it follows
that the channel with the strongest exponential growth will dominate
for $R \to \infty$, while all the others will vanish. This does not
mean, that the partial rates in the physical sense depend on $R$, but
rather, that a correction for the time mixing must be taken into
account.

More precisely: If the channel potentials are \emph{asymptotically}
flat the rate of growth in space for channel $n$ is given
asymptotically as
\begin{equation}
    \label{eq:asymptotic-growth}
    -\Im{\sqrt{2(E_{\mathrm{res}}-E_n)}},
\end{equation}
where $E_{\mathrm{res}}$ is the (complex) eigenvalue of the resonance
in question and $E_n$ is the (real) energy of the corresponding
channel threshold. If this simple exponential behaviour can be shown
in the computed data, then it provides a path to trace the flux of the
ionized electron back to the bound region, from whence it must have
been emitted. As anticipated above, however, a quantitatively correct procedure
to trace back the fluxes would have to take into account to some extent at least the shape of the channel potentials before the asymptotics are reached and the
initial probability distribution. The procedure suggested by equation~(29) of
Ref.~\cite{goldzak_evaluation_2010}, which implies tracing back the fluxes to
the origin $\vecop{r}=\vecop{0}$ with linear velocities, might prove
insufficient in this respect.

In this paper, however, we adopt an even simpler approach. We will see that in
our case, even for rather large decay rates, the correction induced by different
rates of exponential growth in space is in fact so small that we can neglect it
(see Tables~\ref{tab:partial-rates-4-6},~\ref{tab:partial-rates-4-6-e}).

\section{Partial autoionisation rates in 1D
  helium\label{sec:partial_rates_Zee}}

In this section we apply the method developed in
Section~\ref{sec:partial_rates} to the one dimensional helium.

\subsection{General formulation}

After single ionization of the helium atom the remaining ion resembles
a hydrogen atom with different nuclear charge $Z$. For this system the
states are known (see for example
\cite{landau_course_1981,blumel_microwave_1987}) and are given as
\begin{equation}
    \label{eq:hydrogenic-bound-states}
    \phi_{N}(r) = \frac{\sqrt{Z}}{N} \frac{1}{\sqrt{N}}
    \exp\left(-\frac{Zr}{N}\right) \left(\frac{2Zr}{N}\right)
    L_{N-1}^{(1)}\left(\frac{2Zr}{N}\right),
\end{equation}
where $Z$ is the nuclear charge, $N$ is
the principal quantum number of the electron and $r$ its distance from
the nucleus. Using Sturmian functions as defined in
(\ref{eq:1Dsturmians}) this writes as
\begin{equation}
    \label{eq:hydrogenic-bound-states-sturmian}
    \braket{r}{N} := \phi_N(r) \equiv (-1)^N \frac{\sqrt{Z}}{N}
    S_N^{(\frac{N}{Z})}(r).
\end{equation}

We can use the states of the remaining ion to define different decay
channels. Since the outer electron has escaped the system, it is
expected to behave like a Coulomb wave. Furthermore we do not care
about its energy.
Following Sec.~\ref{subsec:numerics} we restrict our
analysis to the region $z_1>z_2$.
Here, $z_1$ is the coordinate of the
outer electron and in accordance with
(\ref{eq:separating-wave-function}) we can define decay
channels by the projection operators
\begin{equation}
    \label{eq:channel-projector}
    P_N = \bbone \otimes \ket{N}\bra{N}.
\end{equation}
In the case of \eZe{} we can execute the projection using the
orthogonality of the Sturmian functions as
\begin{eqnarray}
    \label{eq:eZe-projection}
    P_N\ket{\psi}
    &= \left(\bbone \otimes \ket{N}\bra{N}\right) \sum_{nm} C_{nm}
    \ket{S_n^{(\alpha)}} \otimes \ket{S_m^{(\beta)}}\\
    &= \sum_{nm} C_{nm} \braket{N}{S_m^{(\beta)}}
    \ket{S_n^{(\alpha)}} \otimes \ket{N},
\end{eqnarray}
where $\braket{N}{S_m^{(\beta)}}$ can be calculated as
\begin{eqnarray}
    \braket{N}{S_m^{(\beta)}} &= (-1)^N \frac{\sqrt{Z}}{N}
    \matel{\tilde{S}_N^{(\frac{N}{Z})}}{z_2}{S_m^{(\beta)}},
\end{eqnarray}
which in turn can be computed from (\ref{eq:matel_backrotation}) with
a complex argument, thus reducing the projection to a simple matrix vector
multiplication.

In the \Zee{} case this approach poses a severe challenge:
By introducing perimetric coordinates, we found an efficient basis that
automatically guarantees proper boundary conditions. Unfortunately this
also mixes the coordinates of the two electrons. Therefore, the product
structure of our basis is not the product of two one-electron spaces,
and the found eigenvectors cannot be projected easily. For these
reasons, we resort to projection by integration on a grid in
coordinate space, defining the aforementioned projector by
\begin{equation}
    \label{eq:channel-projector-Zee}
    P_N\psi_{>}(z_1, z_2) :=
    \phi_{N}(z_2) \int_0^{\infty} \phi_{N}(z_2')\psi_{>}(z_1, z_2')
    \mathrm{d}z_2'.
\end{equation}

Following the treatment laid out in Section~\ref{sec:partial_rates}
we choose the volume $V_R$ quadratic as
\begin{equation}
    \label{eq:volume-zee}
    V_R = \{(z_1, z_2) \in \mathbb{R}^2 | 0 \le z_1 \le R; 0 \le z_2
    \le R\}.
\end{equation}
Since the wave function in our case lives completely inside
this region for $R\to\infty$, clearly there can be no flux through the
$z_1$ nor the $z_2$ axis. Consequently, for any fixed, finite $R$, the
flux out of $V_R$ is given by the two contributions of current through
the boundary $\{(R,z_2)|z_2 \le R\}$ and $\{(z_1,R)|z_1 \le R\}$. Due
to the symmetry of the problem they must be identical. Furthermore the
enclosed density must, for the same reason, be distributed equally
among the two triangular parts of $V_R$ defined by the diagonal. Hence
it is enough to look at the triangular region $\{(z_1, z_2) \in
\mathbb{R}^2 | 0 \le z_1 \le R; 0 \le z_2 \le z_1\}$. The flux out of
$V_R$ is given only by the flux through the $z_1=R$ boundary. On this
line, the perpendicular direction is the $z_1$ direction, and
therefore the relevant current density operator is given as
\begin{equation}
    \label{eq:flux-density-operator-z1}
    \tilde{J}(z_1,z_2) = \frac{1}{2}[p_{z_1}\ket{z_1,z_2}\bra{z_1,z_2} +
    \ket{z_1,z_2}\bra{z_1,z_2}p_{z_{1}}].
\end{equation}
Since $\tilde{J}$ only acts in direction of $z_1$, and $P_N$ only in
direction of $z_2$, we have $[P_N,\tilde{J}]=0$. This guarantees the
vanishing of the second term in (\ref{eq:partial-rates}), and
in this sense the choice of $V_R$ is compatible with the given
channels.

Therefore, the relevant component of the partial flux $j_N(z_1, z_2)$
can be calculated exactly like the total flux $j(z_1, z_2)$, with the
projected wave function $\psi_N$ instead of the total wave function
$\psi$:
\begin{equation}
    \label{eq:flux-equation}
    \tilde{j}_{(N)}(z_1, z_2) = \Im\left[\psi_{(N)}^{*}(z_1,
    z_2)\frac{\partial}{\partial z_1}\psi_{(N)}(z_1,
    z_2)\right]
\end{equation}
Thanks to the properties of the Laguerre polynomials, both factors can
be calculated without numeric approximation of the
derivative. Furthermore, since $P_N$ only acts on $z_2$ and the
derivative is in direction of $z_1$, the two operators commute and we
can calculate $\frac{\partial}{\partial z_1}\psi_{N}(z_1,
z_2)$ by projection of $\frac{\partial}{\partial
z_1}\psi(z_1, z_2)$.
In this setup the role of $R$ is assumed by $z_1$. Where no confusion
is possible we will write $z_1$ also for $R$ in the following.

\subsection{Selected resonances}

In this section we demonstrate some results on relatively low lying,
doubly excited states, with both electrons similarly excited. We use
the (4,6) resonance as a prototypical example for the \Zee{}
configuration. For its density and numerical parameters see
Figure~\ref{fig:resonance_4_6}. As stated earlier, the full resonance
wave function is not normalizable. Our numeric approximation, however,
due to the finite extension of our basis, is normalizable. Since a
global factor does not change the equations, we normalize it to
$\lim_{z_1\to\infty}D(z_1) = 1$.

\begin{figure*}
    \includegraphics[width=\textwidth]{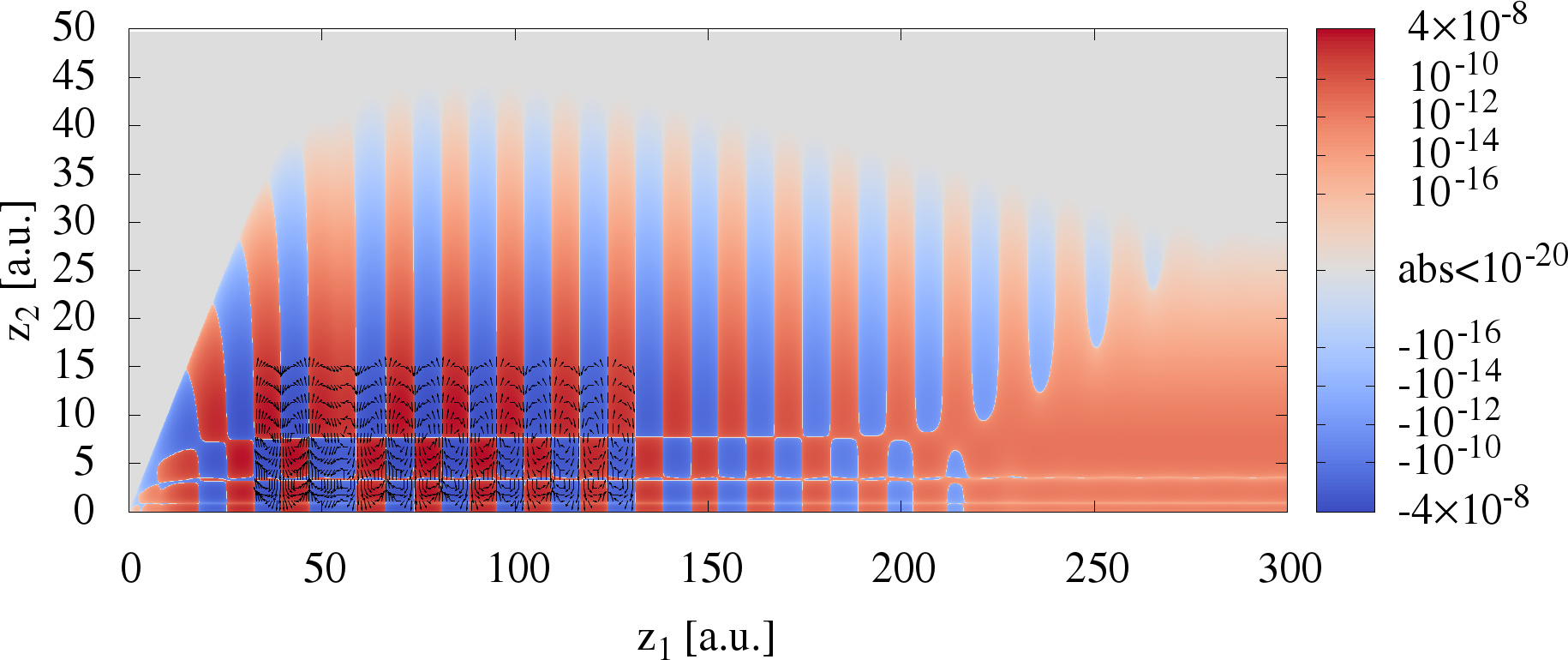}
    \caption{\label{fig:current_density_4_6}
      Current density in direction of $z_1$ for the \Zee{} (4,6)
      resonance. Grey marks all areas with a magnitude smaller than
      $10^{-20}\ a.u$. Positive values (that is flux in positive $z_1$
      direction, \ie outgoing flux) are plotted with a red
      logarithmic color scale, negative values with a blue logarithmic
      color scale. The area shown is the same as for the probability
      density in Figure~\ref{fig:resonance_4_6}, that is the bound
      part and the onset of the asymptotic region. In the bound region
      a checkerboard pattern is observed, which can be attributed to
      circular currents. To illustrate this, part of the associated
      vector field is indicated in the range $z_1 \in
      [30~\mathrm{\au}, 125~\mathrm{\au}]$.
      The slight
      irregularity around $z_1 = 50~\mathrm{\au}$ coincides with the
      vertical nodal line of the probability density at the same position
      (c.f.~Figure~\ref{fig:resonance_4_6}). Values are as
      large as $4 \cdot 10^{-8}~\mathrm{\au}$ in magnitude, with several
      contributions to the integral in
      (\ref{eq:definition-local-quantities}) canceling each other to
      arrive at the much smaller rate $\Gamma = 1.41328 \cdot
      10^{-11}~\mathrm{\au}$
      Observe that, for $z_1 \gtrsim
      150~\mathrm{\au}$, the current, while still oscillating, has
      already decreased to a magnitude which is of the same order as
      in the asymptotic region. Compare this with the noise in
      Figure~\ref{fig:flux_4_6}. The transition region, starting
      around $z_1 \gtrsim 200~\mathrm{\au}$, and the subsequent onset
      of the asymptotic region are shown in more detail in
      Figure~\ref{fig:current_density_zoom_4_6}.}
\end{figure*}

\begin{figure*}
    \includegraphics[width=\textwidth]{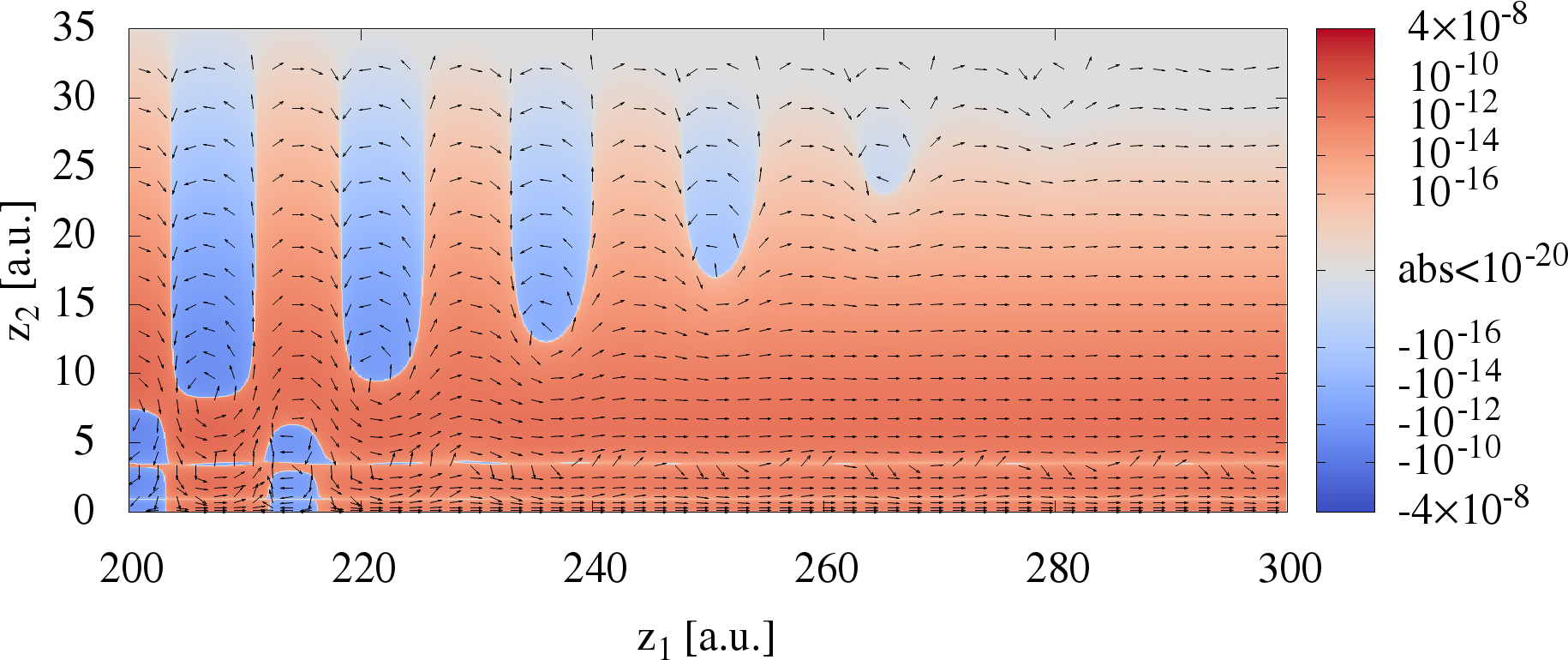}
    \caption{\label{fig:current_density_zoom_4_6}
      Current density in direction of $z_1$ for the \Zee{} (4,6)
      resonance (color code) and direction of current (arrows). Shown
      is a closeup of the transition region from the bound to the
      asymptotic region. On the left, around $z_1 \approx 200~\mathrm{\au}$,
      the current behaviour as observed in the checkerboard region of
      Figure~\ref{fig:current_density_4_6} can be seen. Strong
      circular currents fill that area. As $z_1$ increases the
      circular currents disappear and a mostly parallel outward
      current prevails.}
\end{figure*}

The current density that has to be integrated is shown in
Figures~\ref{fig:current_density_4_6}~and~\ref{fig:current_density_zoom_4_6}. The
strong circular currents in the bound region necessitate a
particularly careful integration. This is carried out with the
adaptive Gauss-Kronrod method as provided by the GNU Scientific
Library. Note that the current density in the asymptotic region is
much better behaved and consequently easier to integrate.

\FloatBarrier

\begin{figure}
    \includegraphics[width=0.68\textwidth]{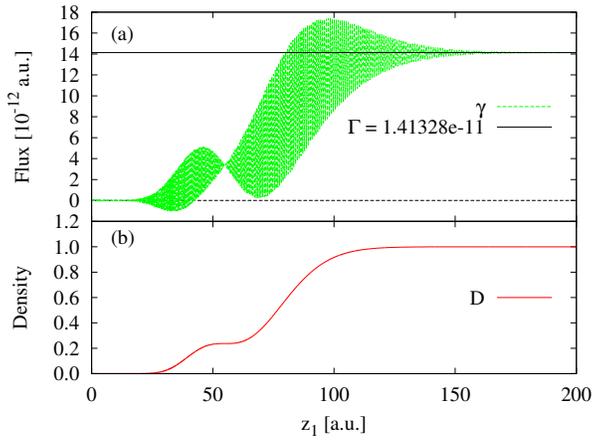}
    \caption{\label{fig:flux_4_6}
      Total current in direction of $z_1$ (a) and integrated
      density (b) for the \Zee{} (4,6) resonance
      (c.f. Figure~\ref{fig:resonance_4_6}).}
\end{figure}

\begin{figure}
    \includegraphics[width=0.68\textwidth]{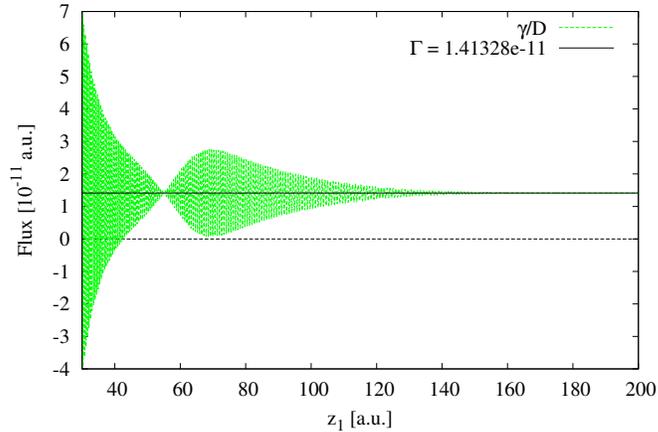}
    \caption{\label{fig:total-flux-bound-normalized}
      Total current, normalized by the integrated density for the
      \Zee{} (4,6) resonance. On
      average it agrees with the rate as given by the eigenvalue
      calculation even in the bound part. Very close to the origin
      both numerator and denominator become small, leading to
      numerical problems.}
\end{figure}

Figure~\ref{fig:flux_4_6} shows the resulting integrated quantities
$\gamma$ and $D$,
(\ref{eq:local-partial-rates}, \ref{eq:definition-local-quantities}),
for the (4,6) resonance. $D$ quickly reaches 1. This
is because the bulk of the density is in the bound region, where the
asymptotic regions only contribute very small amounts (cf
Fig.~\ref{fig:resonance_4_6}). We see that $\gamma$ follows the shape
of $D$. Figure~\ref{fig:total-flux-bound-normalized} shows the
normalized current with the rate as calculated from the eigenvalue
superimposed. Even deep into the bound part the calculations
agree. Only very close to the origin numerical problems dominate.

\FloatBarrier

\begin{figure}
    \includegraphics[width=0.68\textwidth]{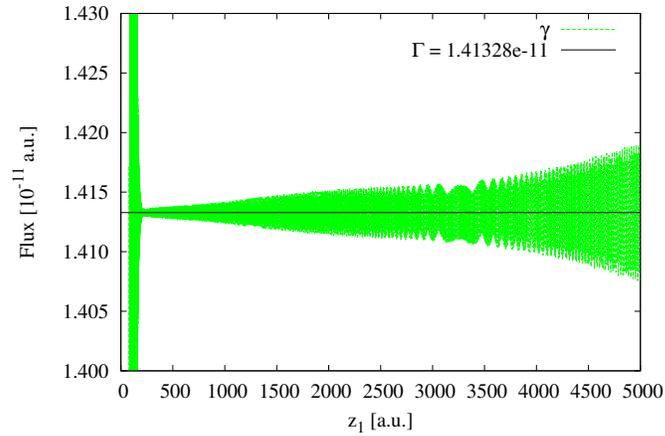}
    \caption{\label{fig:total-flux-asymptotic}
      Total current in the asymptotic region for the
      \Zee{} (4,6) resonance. The average value
      agrees with the rate known from the eigenvalue calculation over
      a wide region of configuration space. Even at large distances
      from the nucleus the noise level is below 1 per cent.}
\end{figure}

Let us take a look at the total current in the asymptotic
region. This is shown in Figure~\ref{fig:total-flux-asymptotic}. While
far away from the nucleus also numerical problems increase, the
current calculation agrees with the rate from the eigenvalue
calculation for a wide range of configuration space.

\FloatBarrier

\begin{figure}
    \includegraphics[width=0.68\textwidth]{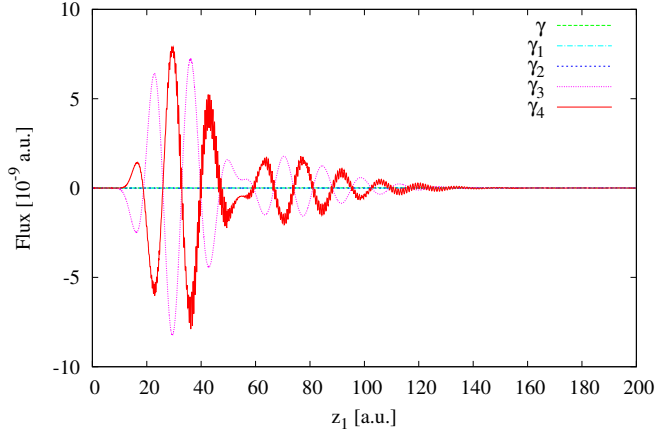}
    \caption{\label{fig:partial-flux-bound}
      Partial currents of the \Zee{} (4,6) resonance for all open
      channels in the bound region of configuration space. The two
      most dominant contributions are seen to cancel each other
      despite significant oscillations. The interplay of the partial
      rates indicates that the channels are not yet separated.}
\end{figure}

We now turn to the analysis of partial
rates. Figure~\ref{fig:partial-flux-bound} shows the partial currents
in all energetically possible channels in the bound region. We see
how the two dominant channels interact. Their amplitude is three
orders of magnitude larger than the total current, to which they add
up. Here we do not show the normalized currents $\gamma/D$, but their
unnormalized counterparts. This is done to avoid numerical problems
and does not alter the analysis, in that it would, at every point, only
add a factor that is constant for all channels.

\begin{figure}
    \includegraphics[width=0.68\textwidth]{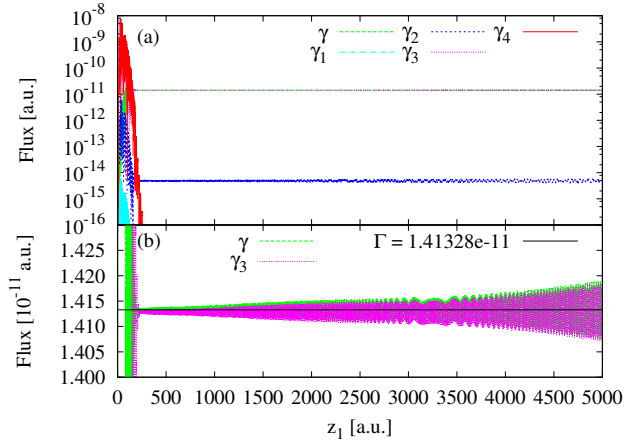}
    \caption{\label{fig:partial-flux-asymptotic}
      Partial currents of the \Zee{} (4,6) resonance for (a) all open
      channels in the asymptotic
      regime, (b) the dominant channel. In the logarithmic plot (a) we
      see $\gamma_1$ and $\gamma_4$ quickly drop to values which, due
      to the finite precision arithmetics, need to be considered as
      numerically zero. The dominant contribution to the total current
      stems from the $\gamma_3$ component, with a small contribution from
      $\gamma_2$. The linear plot (b) confirms this.}
\end{figure}
Figure~\ref{fig:partial-flux-asymptotic} shows the asymptotic
behaviour of the partial currents. Channels $N=1,4$ quickly drop to
numerically zero. Channels $N=2,3$ become constant and clearly
separate. Indeed this justifies the notion of an asymptotic region
directly from the data without an ad hoc definition.

The $N=3$ channel is the dominant one. The only other contribution is
from the $N=2$ channel. We estimate partial rates as the average
values of the partial currents in the interval $z_1 \in
[200~\mathrm{\au}, 5000~\mathrm{\au}]$. The results are shown in
Table~\ref{tab:partial-rates-4-6}.
Note, in particular, the high accuracy of our approach, which is
reflected in the agreement of the sum of partial rates with the
total rate in 6 significant digits.
Clearly in this case the correction for the exponential growth of the
partial currents that was discussed in
Section~\ref{subsec:partial_rates} is not necessary. This could be due
to the rather small rates. Next we examine a case with much stronger
decay.
\begin{table}
    \caption{\label{tab:partial-rates-4-6}
      Comparison of the rates for the (4,6) state of \Zee{} helium as
      obtained from eigenvalue
      ($\Gamma$) and current calculations. The latter are calculated
      as the averages over part of the asymptotic region which, from the
      current plots (see Figure~\ref{fig:partial-flux-asymptotic}), is
      determined as the interval $z_1 \in [200~\mathrm{\au},
      5000~\mathrm{\au}]$ .}
    \begin{tabular}{ccc}
        Source&Rate&Standard deviation\\\hline
        $\Gamma$&$1.41328 \cdot 10^{-11}~\mathrm{\au}$&\\
        $\gamma$&$1.41328 \cdot 10^{-11}~\mathrm{\au}$&$1.4 \cdot
        10^{-14}~\mathrm{\au}$\\
        $\gamma_3$&$1.41279 \cdot 10^{-11}~\mathrm{\au}$&$1.4 \cdot
        10^{-14}~\mathrm{\au}$\\
        $\gamma_2$&$0.00049 \cdot 10^{-11}~\mathrm{\au}$&$3.0 \cdot
        10^{-16}~\mathrm{\au}$\\
        $\gamma_3+\gamma_2$&$1.41328 \cdot 10^{-11}~\mathrm{\au}$&$1.5 \cdot
        10^{-14}~\mathrm{\au}$
    \end{tabular}
\end{table}

\FloatBarrier

\begin{figure}
    \includegraphics[width=0.68\textwidth]{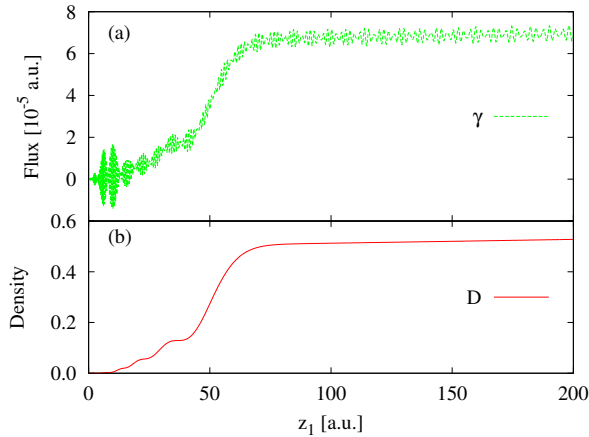}
    \caption{\label{fig:total-flux-4-7-e-bound}
      Total current in direction of $z_1$ (a) and integrated
      density (b) for the \eZe{} (4,7,e) resonance
      (c.f. Figure~\ref{fig:resonance_4_7_e}).}
\end{figure}

\begin{figure}
    \includegraphics[width=0.68\textwidth]{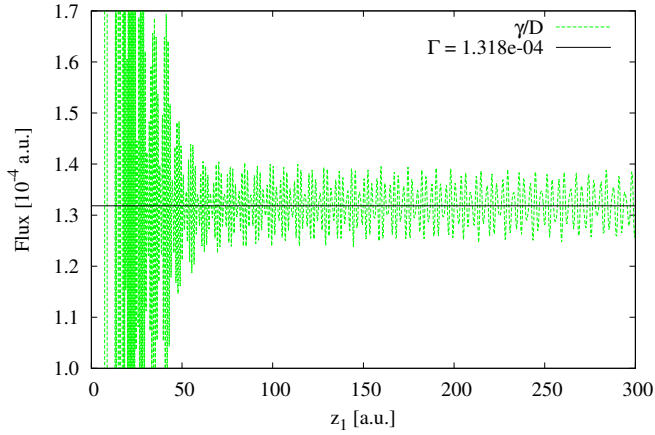}
    \caption{\label{fig:total-flux-4-7-e-normalized}
      Total current, normalized by the integrated density for the
      \eZe{} (4,7,e) resonance. On
      average it agrees with the rate as given by the eigenvalue
      calculation even in the bound part. Close to the origin
      both numerator and denominator become small, leading to
      numerical problems.}
\end{figure}

\begin{figure}
    \includegraphics[width=0.68\textwidth]{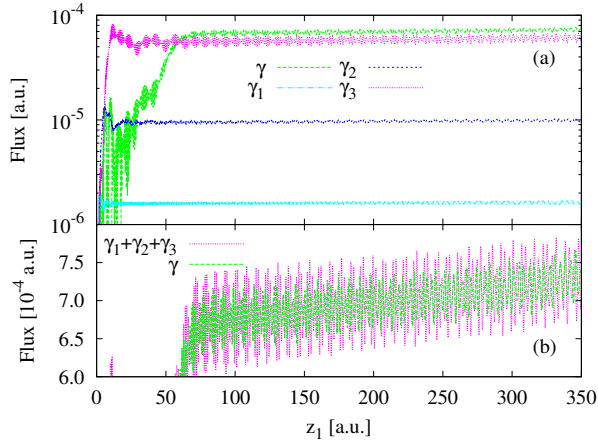}
    \caption{\label{fig:total-flux-4-7-e-asymptotic}
      Partial currents of the \eZe{} (4,7,e) resonance for (a) all
      contributing channels in the asymptotic
      regime, (b) the sum of all contributing channels compared to the
      total current. In the logarithmic plot (a) we
      see that three channels, $\gamma_1$, $\gamma_2$, and $\gamma_3$,
      contribute with increasing relevance. The linear plot (b) shows
      the agreement of the sum of partial currents and total current
      and the expected growth with increasing $z_1$. This is corrected
      for by the integrated density
      (c.f. Figure~\ref{fig:total-flux-4-7-e-asymptotic-normalized}).}
\end{figure}

We now present results for the \eZe{} configuration using the
$(4,7,\mathrm{even})$ resonance as an example. Its probability density
has already been shown in
Figure~\ref{fig:resonance_4_7_e}. Figure~\ref{fig:total-flux-4-7-e-bound}
shows the total current prior to normalization together with the
integrated density.
Both exhibit the expected structured increase in
the bound region with $z_1<100~\mathrm{\au}$ Contrary to the \Zee{}
case (cf. Figure~\ref{fig:flux_4_6}) the asymptotic region is not
flat, but is in fact increasing exponentially. We therefore make no
attempt to normalize the state such that the integrated density is
1. Consequently all not normalized currents are practically in
arbitrary units. Figure~\ref{fig:total-flux-4-7-e-normalized}
demonstrates, that the normalized total current again agrees with the
decay rate as calculated from the eigenvalue.
The exponential increases of current and density cancel and give a
proper constant rate, save for some variation introduced by numerics.
We now turn to the analysis of partial rates. The partial currents
without normalization are shown in
Figure~\ref{fig:total-flux-4-7-e-asymptotic}.
In part (a) we see that all three open channels have partial currents
of non negligible magnitude. Part (b) shows that their sum indeed
reflects the total current. This allows us to understand the ripples
that have been shown in the probability density of the state
(Figure~\ref{fig:resonance_4_7_e}): The shape of the outgoing part in
ionic channel $N$ is characterized by $N$ antinodes in $z_2$
direction. Hence the antinode closest to the $z_1$ axis is occupied
by all three outgoing channels, the second one by the $N=2,3$ channels
and the third one only by the $N=3$ channel. In $z_1$ direction the
wave functions resemble Coulomb waves with different momenta $k$ as
determined by the differences in energy between the corresponding
ionic state of the remaining system and the initial energy of the
resonance. These different momenta produce a beating effect that shows
up as ripples in the density.

\begin{figure}
    \includegraphics[width=0.68\textwidth]{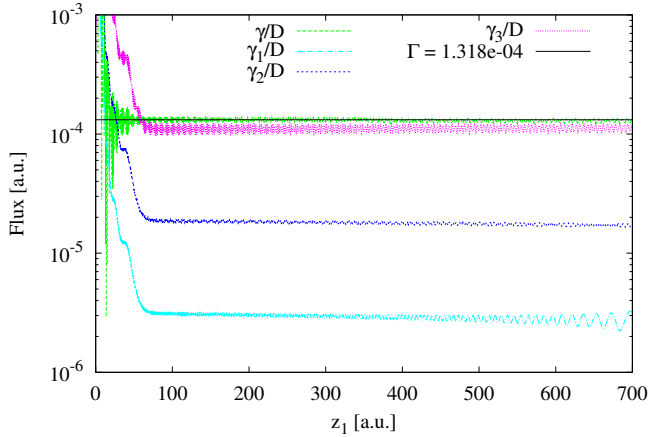}
    \caption{\label{fig:total-flux-4-7-e-asymptotic-normalized}
      Partial currents of the \eZe{} (4,7,e) resonance, normalized by
      the integrated density. See
      Figure~\ref{fig:total-flux-4-7-e-asymptotic} for the agreement
      of partial and total currents. The total rate as calculated from
      the eigenvalue is overlaid. While the curves are not constant,
      even on average in the asymptotic regime, they change so little
      as to permit the calculation of a meaningful average (cf
      discussion at the end of
      Section~\ref{subsec:partial_rates}). The results are shown in
      Table~\ref{tab:partial-rates-4-6-e}.}
\end{figure}

For a quantitative analysis of the partial currents we turn to
Figure~\ref{fig:total-flux-4-7-e-asymptotic-normalized}.
It contains the normalized partial and total currents together with
the total rate as calculated from the eigenvalue. From this we fit
constants for the partial rates in the asymptotic region, here taken
as $z_1 \in [100~\mathrm{\au},700~\mathrm{\au}]$. The results are
shown in Table~\ref{tab:partial-rates-4-6-e}. Even with large decay
rates and several contributing channels as in this case, the partial
rates agree with the total rate. Moreover, despite the noticeable
exponential growth, in a large enough region of configuration space
that still certainly can be called asymptotic, they are
approximately constant.

\begin{table}
    \caption{\label{tab:partial-rates-4-6-e}
      Comparison of the rates as obtained from eigenvalue
      ($\Gamma$) and current calculations. The latter are calculated
      as the averages over part of the asymptotic region which, from the
      current plots (see Figure~\ref{fig:partial-flux-asymptotic}), is
      determined as the interval $z_1 \in [100~\mathrm{\au},
      700~\mathrm{\au}]$ .}
    \begin{tabular}{ccc}
        Source&Rate&Standard deviation\\\hline
        $\Gamma$&$1.3183 \cdot 10^{-4}~\mathrm{\au}$&\\
        $\gamma$&$1.3202 \cdot 10^{-4}~\mathrm{\au}$&$1.4 \cdot
        10^{-7}~\mathrm{\au}$\\
        $\gamma_3$&$1.1109 \cdot 10^{-4}~\mathrm{\au}$&$1.5 \cdot
        10^{-7}~\mathrm{\au}$\\
        $\gamma_2$&$0.17970 \cdot 10^{-4}~\mathrm{\au}$&$1.3 \cdot
        10^{-8}~\mathrm{\au}$\\
        $\gamma_1$&$0.02928 \cdot 10^{-4}~\mathrm{\au}$&$3 \cdot
        10^{-9}~\mathrm{\au}$\\
        $\gamma_3+\gamma_2+\gamma_1$&$1.3198 \cdot 10^{-4}~\mathrm{\au}$&$1.7 \cdot
        10^{-7}~\mathrm{\au}$
    \end{tabular}
\end{table}

\section{Conclusion\label{sec:conclusion}}

We have presented a theoretical framework for the study of partial
decay rates. Partial decay channels can be resolved on the basis of
projection operators of wave functions. This was combined with the
method of complex rotation to allow for high precision analysis of
resonances. For the first time, partial rates have been calculated
without the assumption of a specific final state from complex
rotation. The utility of the method has been demonstrated by its
application to two one-dimensional helium model systems. Our approach
was successful, despite the radically different rates, for both the
extremely long-lived \Zee{} configuration and the \eZe{} case, which
exhibits strong decay. Notwithstanding the long-range nature of the
involved Coulomb potential, an asymptotic regime was clearly observed
for both situations. The approach can be applied to other decay
problems like higher dimensional helium. Here, building
on earlier work in which backrotated wave functions have already been
calculated \cite{madronero_ab_2008}, what remains to be done is the
definition of proper channel projectors. This endeavour seems to be
promising in the light of the total rates that have been compared for 1D,
2D, and 3D helium \cite[Figs.~2,3]{madronero_decay_2005} and that were
found to be on the same order for 2D, 3D and 1D \eZe{} rates, while
the 1D \Zee{} rates are much smaller as also demonstrated here.

% Specify following sections are appendices. Use \appendix* if there
% only one appendix.
\appendix
%\section{}

% If you have acknowledgments, this puts in the proper section head.
%\begin{acknowledgments}
% put your acknowledgments here.
\section*{Acknowledgements}
We acknowledge the use of the computing resources provided by the
Black Forest Grid Initiative, by bwGRiD (http://www.bw-grid.de), and
the computing time granted by the John von Neumann Institute for
Computing (NIC), and alotted on the supercomputer JUROPA at J\"ulich
Supercomputing Centre (JSC). Furthermore, funding by DFG Research Unit
760 is gratefully acknowledged. P.~L. acknowledges financial support
from the Humboldt Foundation through Fellowship No. 1139948.
%\end{acknowledgments}

%\section*{References}
% Create the reference section using BibTeX:
%\bibliographystyle{unsrtnat} % unsrt.bst: numerical (Vancouver) style
%\bibliographystyle{plain} % plain.bst: natbib plain
%\bibliographystyle{abbrv} % abbrv.bst: natbib abbrv
%\bibliographystyle{apsrev} % apsrev.bst
\bibliographystyle{myiopart-num}
\bibliography{2012_partial_rates}

\end{document}